# Convergent beam electron holography for analysis of van der Waals heterostructures

*Tatiana Latychevskaia[1], Colin Robert Woods[2,3], Yi Bo Wang[2,3], Matthew Holwill[2,3], Eric Prestat[4], Sarah J. Haigh[2,4], Kostya S. Novoselov[2,3]*

[1]*Institute of Physics, Laboratory for ultrafast microscopy and electron scattering (LUMES), École Polytechnique Fédérale de Lausanne (EPFL), CH-1015 Lausanne, Switzerland*

[2]*National Graphene Institute, University of Manchester, Oxford Road, Manchester, M13 9PL, UK*

[3]*School of Physics and Astronomy, University of Manchester, Oxford Road, Manchester, M13 9PL, UK*

[4]*School of Materials, University of Manchester, Oxford Road, Manchester, M13 9PL, UK*

**Corresponding author:**

Tatiana Latychevskaia

Institute of Physics, Laboratory for ultrafast microscopy and electron scattering (LUMES), École Polytechnique Fédérale de Lausanne (EPFL), CH-1015 Lausanne, Switzerland

Tel: +41 44 635 6046

Mail: tatiana@physik.uzh.ch

**Corresponding author:**

Kostya S. Novoselov

School of Materials, University of Manchester, Oxford Road, Manchester, M13 9PL, UK

Tel: +44(0)161 275-4119

Mail: konstantin.novoselov@manchester.ac.uk

**ABSTRACT**

Van der Waals heterostructures, which explore the synergetic properties of two-dimensional (2D) materials when assembled into three-dimensional stacks, have already brought to life a number of exciting new phenomena and novel electronic devices. Still, the interaction between the layers in such assembly, possible surface reconstruction, intrinsic and extrinsic defects are very difficult to characterise by any method, because of the single-atomic nature of the crystals involved. Here we present a convergent beam electron holographic technique which allows imaging of the stacking order in such heterostructures. Based on the interference of electron waves scattered on different crystals in the stack, this approach allows one to reconstruct the relative rotation, stretching, out-of-plane corrugation of the layers with atomic precision. Being holographic in nature, our approach allows extraction of quantitative information about the three-dimensional structure of the typical defects from a single image covering thousands of square nanometres. Furthermore, qualitative information about the defects in the stack can be extracted from the convergent diffraction patterns even without reconstruction – simply by comparing the patterns in different diffraction spots. We expect that convergent beam electron holography will be widely used to study the properties of van der Waals heterostructures.

**SIGNIFICANCE STATEMENT**

Assembling 2D materials into vertically stacked heterostructures allows an unprecedented control over their properties. The interaction between the individual crystals plays the crucial role here, thus, the information about the local atomic stacking is of great importance. Still, there are no techniques which would allow investigation of the stacking between such crystals with any reasonable throughput. We present the use of convergent beam electron diffraction (CBED) to investigate the quality of the interface in such heterostructures. We demonstrate that defects such as misorientation, strain, ripples, and others can be visualized and quantitative information about such structures can be easily extracted. Furthermore, CBED images can be treated as holograms, thus their reconstruction gives three-dimensional profiles of the heterostructures over a large area.

**MAIN TEXT**

Stacking 2D materials into van der Waals heterostructures offers an unprecedented control over the attributes of the resulting devices (1, 2). Initially the individual layers in the stack were considered to be independent, which, offers a reasonable zero-order approximation of the properties of such heterostructures. However, as we gain better and better control over the stacking arrangement between the individual components and the cleanliness of the interfaces – the interaction between the individual crystals becomes more and more important, and can even dominate the performance of the structures.

Still, it is very difficult to extract the detailed information about the stacking. Cross-sectional transmission electron microscopy (TEM) imaging (3) allows atomic scale information on the structure and chemistry of the buried interface between the individual crystals to be obtained. Unfortunately, this technique requires sophisticated sample preparation, is time consuming, and only yields data on a thin slice of the sample, which is not necessarily representative of the large area device. Preferential scattering detection in the scanning TEM has been recently demonstrated to allow determination of atomic stacking for well-aligned graphene/boron nitride heterostructures, but it requires a custom aperture configuration and is based on atomic resolution imaging compared with relaxed density functional theory (DFT) modelling so is inherently limited to a small field of view (4). Thus, the whole field of van der Waals heterostructures would benefit enormously from a technique which allows one to extract three-dimensional structure for the buried interfaces inside such stacks on a larger scale.

Convergent beam electron diffraction (CBED) (5) has been previously applied to three-dimensional crystals where it provides a valuable method for studying crystallographic structure (6-9), and measurements of strain (10, 11) and specimen thickness (12, 13) for a nanoscale volume. The choice of the electron beam convergence angle (14, 15), defocusing distance, lens aberrations and specimen thickness allow precise control of the volume of material analysed in a single measurement. However, accurate interpretation of a general CBED pattern is not straightforward, requiring careful comparison to simulated structures, which often limits application of the technique. CBED on thin van der Waals heterostructures (1, 2) would deliver a dramatically larger amount of information, which is intuitively easy to interpret and immediately results in both qualitative and quantitative data about the stack. Furthermore, CBED of van der Waals heterostructures can be considered as a hologram, so conventional holographic reconstruction techniques can be applied, thus delivering information about the three-dimensional arrangement between the crystals in the stack (including the local strain, lattice orientations, local vertical separation between the layers, *etc.*) which is not accessible by conventional TEM imaging (16, 17).

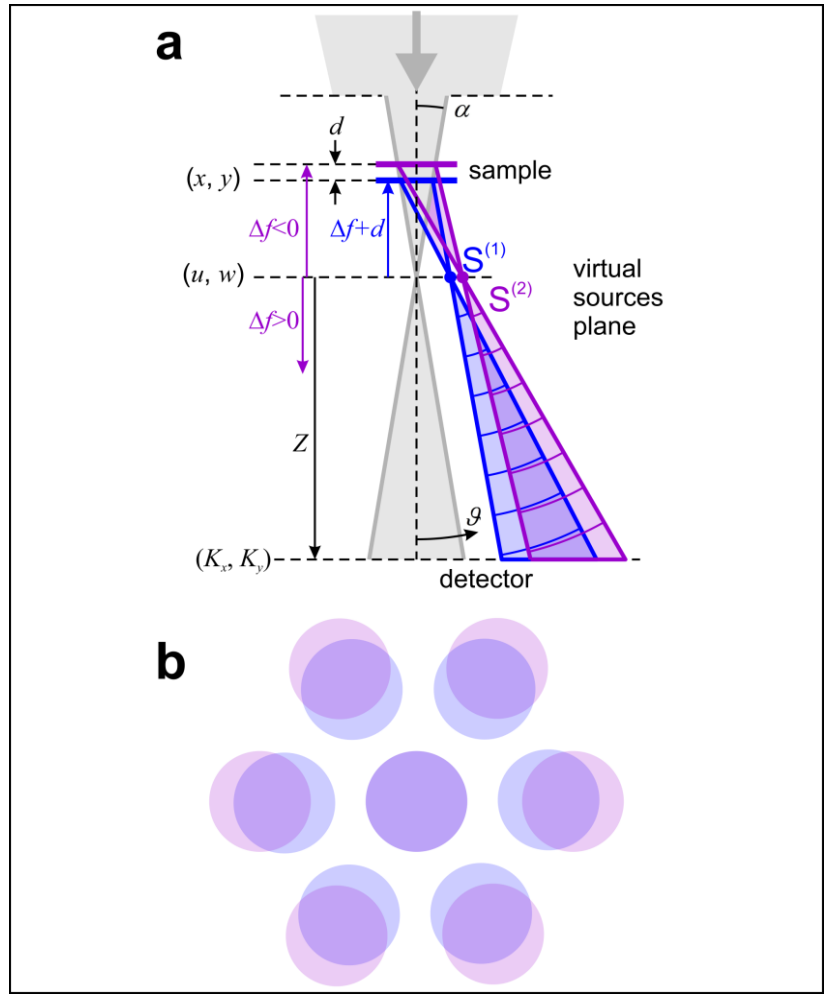

Fig. 1. Experimental arrangement for convergent beam electron diffraction. (a) Schematics of CBED on a bilayer system. Here $\Delta f$ is the sample $z$-position counted from the focus of the electron beam (in this particular case underfocus $\Delta f<0$ CBED conditions are shown), $Z$ is the distance from the virtual sources plane to the detector, $S^{(1)}$ and $S^{(2)}$ are the virtual sources for the first-order CBED spots of bottom (1) and top (2) crystals in the heterostructure stack respectively. $\vartheta$ is the angular coordinate on the detector. (b) Distribution of CBED spots on a detector in the case of two aligned crystals with slightly different lattice constants.

A CBED pattern from a single layer of graphene or hexagonal boron nitride (hBN) consists of finite-sized spots arranged into a six-fold symmetrical pattern. The centres of the spots have the same positions as the diffraction peaks, given by $\sin\vartheta=\lambda/a$, where $\lambda$ is the wavelength, $\vartheta$ is the diffraction angle and $a$ is the period of the crystallographic planes. The size of a CBED spot on the detector depends on the convergent angle and the diameter of the limiting aperture and it remains the same independent of the z-position of the sample (or defocusing distance $\Delta f$). The sample area imaged within a CBED spot corresponds to the illuminated area, whose diameter can be approximated as $2\Delta f \tan\alpha$, where $\alpha$ is the convergence semi angle of the electron beam, Fig. 1a.

Probing the samples with a convergent (underfocus, $\Delta f < 0$) or divergent (overfocus, $\Delta f > 0$) incident wavefront is achieved by changing the $z$ position of the sample.

In case of CBED on a bilayer structure (for instance graphene/hBN stack – the type of devices we concentrate on in this paper), the electron beams diffracted on each layer interfere at the positions where the CBED patterns of individual layers overlap (Fig. 1b), thus creating a specific interference pattern (Fig. 2). Such interference patterns contain information about the local interatomic spacing (local strain), the vertical distance between the crystal layers, the relative orientation between the layers, etc.

## Results

**Simulated CBED of perfect crystals.** Fig. 2 demonstrates modelling of CBED for several typical bilayer heterostructures consisting of perfect crystals (the simulation procedure is described in Supporting Information). The interference pattern in a CBED spot can be interpreted as being created by a superposition of two waves originating from two corresponding virtual sources, as sketched in Fig. 1a. If the two stacked crystals have the same lattice constant and the same orientation (as, for instance, in the case of AA or AB stacked bilayer graphene) then the virtual spots are found almost at the same position and no interference pattern is observed in CBED spots. A relative rotation between the layers (Fig. 2a – d), or a slight mismatch of the lattice constants (Fig. 2e – h), lead to the virtual sources, and correspondingly, the CBED spots, to appear at slightly different positions, Fig. 1b. As a result, interference between the CBED spots occurs, and a fringed interference pattern is observed where the spots overlap. If the two graphene crystals in the double layer are rotated with respect to each other by a small angle $\beta$, the CBED patterns from the two crystals will be rotated relative to each other by the same angle. Diffracted electron waves, which originate from the separate layers, but arrive to the same point in the CBED detector plane, gain different phases, and the difference is now proportional to the rotation between the layers. As a result, radially distributed interference fringes will be observed, with the period of fringes within a particular CBED spot, dependent on the angle of the rotation between the layers, Fig. 2a – d.

In the case of two crystals with aligned crystallographic directions but with slightly different interatomic spacing (as, for instance, the case for graphene and hBN) – the CBED spots will be shifted in the radial direction, Fig. 1b. hBN has a 1.8% larger basal plane lattice spacing than graphene and this will result in the appearance of an interference patterns with tangentially distributed fringes, Fig. 2e – h. The period and the tilt of the fringes are unambiguously defined by the lattices' periods, the probing electron beam wavelength, the $z$-position of the sample, the relative rotation between the layers, and the distance between the layers (as shown in Supporting Information). Interestingly, CBED patterns of such bilayer samples are extremely sensitive to the atomic arrangements in the layers. If the local stacking under the centre of the electron beam is AA (Fig. 2e) – the pattern of interference fringes has maxima at the centre of the CBED spots, Fig. 2f. If the local atomic stacking is AB (Fig. 2g) – the pattern of interference fringes in the CBED spots is shifted and as a result, the intensity distribution in the opposite CBED spots (such as (1-100) and -(1100)) are not mirror-symmetric, Fig. 2h.

Also the distance between the layers affects the interference pattern. The phase difference between the electron waves diffracted from the two identical and aligned crystals separated by a distance $\Delta z$ is given by

$$\Delta \varphi_z = \Delta z \frac{2\pi}{\lambda} (1 - \cos \vartheta), \qquad (1)$$

where $\lambda$ is the electron wavelength (4.2 pm for 80 keV electrons). Since $\sin \vartheta = \frac{\lambda}{a}$, the phase shift within the CBED spot can be approximated as:

$$\Delta \varphi_z \approx \frac{\pi \lambda \Delta z}{a^2}, \qquad (2)$$

where $a$ is the lattice period which defines the order of the particular CBED spot. It is clear that for a bilayer graphene ($a = 2.13$ Å for the first-order CBED spots), with a typical interlayer distance of about 3.35 Å, the phase difference is negligible (about 0.1 rad) and remains almost constant over the entire CBED spot and thus no interference fringes should be observed within the CBED spots (as indeed has been reported previously (18)). Further discussion on the period of the interference fringes and examples of the reconstruction of the exact interlayer distances from CBED patterns are provided in Supporting Information.

We should stress that even in the case of a monolayer, the intensity distribution within each CBED spot is not an image of the sample but a far-field distribution of the intensity of the wave scattered by the sample, which in fact is an inline hologram of the sample area.

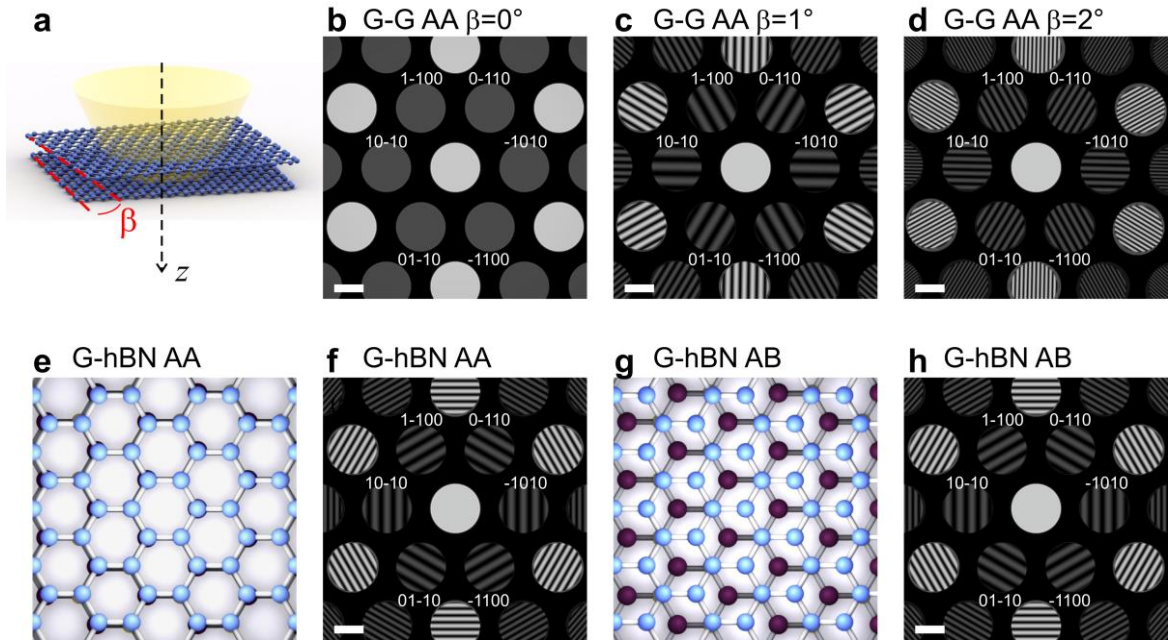

Fig. 2 Simulated CBED patterns for various 2D bilayer heterostructures. (a) Illustration of a bilayer graphene illuminated by a convergent electron beam with one graphene layer rotated by an angle $\beta$. (b) – (d) Simulated CBED patterns of a double layer graphene for different $\beta$. (e) Sketch of the top view of a graphene-hBN stack with AA atomic stacking in the centre of the beam, and (f) the simulated CBED pattern. (g) Sketch of the top view of graphene-hBN with AB stacking in the centre of the beam, and (h) the simulated CBED pattern. For these simulations the distance between the layers is 3.35 Å, $\Delta f$=-3 µm and the imaged area is about 50 nm in diameter. All scale bars correspond to 2 nm$^{-1}$.

**Simulated CBED for crystals with out-of-plane atomic displacements.** To quantitatively study the relation between the atomic defects and the CBED pattern we simulated out-of-plane and in-plane displacements of atoms in the graphene lattice and the corresponding phase distributions in the detector (far-field) plane, shown in Fig. 3. When atoms are displaced out of the crystal plane by a distance $\Delta z$, the scattered wave gains additional phase shift given by Eq. 1. Because $\Delta \varphi_z$ is an even function of $\vartheta$, the added phase shift causes equal intensity change (increase or decrease of the intensity) in opposite CBED spots (see Supporting Information for the derivation of the formulae for the phase shift). A simulated CBED pattern of an out-of-plane defect (bubble, Fig. 3a) with a maximum height in the centre of 2 nm is presented in Fig. 3b, with the phase distribution in the far-field is shown in Fig. 3c. Equation 2 allows estimation of the height of the bubble from the measured phase with 90% precision. The discrepancy is because of the diffraction effects (which are strong for 4 nm wide bubble) due to perfect coherence of the electron beam assumed in the simulations. This Fresnel diffraction effect can be compensated and the true shape of the bubble restored by performing deconvolution with the free-space propagator as explained in detail in Supporting Information. Such diffraction effects are much less pronounced in the experimental images where electron waves are only partially coherent (Supporting Information, section "CBED pattern of an edge"), and where the use of formula (2) gives excellent level of accuracy.

A CBED pattern of a monolayer contains only the amplitude information of the wave, losing the phase part of the signal. Consequently a direct recalculation of the CBED pattern into the atomic distribution is impossible. However, the CBED pattern intensity distribution unambiguously relates to the atomic three-dimensional displacements, and the atomic displacements in principle can be recovered by simulating a matching CBED pattern of a lattice with modelled displacements. The situation is considerably improved in the case of CBED of a bilayer system. Here, the second layer adds a second wavefront that acts as a reference wave. This situation can be considered as a form of an off-axis holography, where the wavefront, scattered on one of the crystals, is treated as the object wave, and the wavefront scattered on the other layer – as the reference wave. The resulting interference pattern forms an off-axis hologram, which can be reconstructed to give the amplitude and phase distributions of the wavefront at the position of each CBED spot. The reconstruction approach which we use here is based on an established procedure for reconstruction of off-axis holograms (19-21), and involves two Fourier transforms (22) (more details are provided in Supporting Information). From the set of phase distributions recovered for individual CBED spots, the atomic displacements in the layers can be immediately recovered.

Fig. 3d shows a sketch of a bilayer graphene/hBN heterostructure with out-of-plane displacement of atoms in the graphene layer. The corresponding simulated CBED pattern is shown in Fig. 3e. An out-of-plane atomic displacement results in an additional phase shift, which is the same for all CBED spots of the same order. Thus, in order to extract the out-of-plane atomic displacements, the symmetric component of the CBED signal is extracted by averaging the phase distribution from all CBED spots of the same order, producing $\Delta\varphi_z$. $\Delta z$ is calculated from $\Delta\varphi_z$ by applying Eq. 2, Fig. 3f.

**Simulated CBED of crystals with in-plane atomic displacement.** When atoms are displaced within the crystal plane by a distance $\Delta x$ the additional phase shift is given by

$$\Delta\varphi_x = -K_x \Delta x, \qquad (3)$$

where $(K_x, K_y)$ are the coordinates in the far-field (detector) plane. Fig. 3g shows a sketch of a graphene layer with an in-plane displacement $\Delta x = -10$ pm for the atoms in the positive semiplane ($x > 0$). The corresponding simulated CBED pattern is shown in Fig. 3h. Because the phase shift $\Delta\varphi_x$ is an odd function of $K_x$, an in-plane displacement give rise to opposite intensity variation in opposite CBED spots, as shown in Fig. 3h. As follows from Eq. 3, for $\Delta x = 10$ pm, the maximal phase shift in the first-order CBED spots amounts to ±0.3 radian, which is confirmed by the phase distribution shown in Fig. 3i. In Fig. 3i, vertical interference fringes observed in the centre of each CBED spots are due to diffraction on a knife-edge. Because infinite coherence is assumed in the simulation, Fresnel fringes appear in the region of about $2.4\sqrt{\lambda \Delta f/2}$. Again, if we perform deconvolution with the free-space propagator (see Supporting Information) this Fresnel diffraction effect can be completely compensated. Such diffraction effects should be much less pronounced in experimental images where electron waves are only partially coherent.

According to Eq. 3, an in-plane displacement does not cause intensity variations (irregularities of fringes) in CBED spots that are orthogonal to the direction of the displacement (note that there is no change in the intensity distributions in (1-210) and (-12-10) spots, Fig. 3h,i). In order to quantitatively reconstruct the in-plane atomic shifts in the $x$-direction, the difference between the phase distributions in the opposite CBED spots along the $K_x$-direction should be calculated and divided by 2, thus extracting only the antisymmetric component of the signal. $\Delta x$ is then calculated from $\Delta \varphi_x$ by applying Eq. 3.

Fig. 3j shows a sketch of a graphene/hBN heterostructure with atomic in-plane displacement of some of graphene atoms (the whole right semiplane is shifted by 10 pm). The corresponding simulated CBED pattern is shown in Fig. 3k. The phase distributions were reconstructed for each CBED spot, and $\Delta x$ was obtained from the reconstructed phase distributions as described above. Fig. 3l shows the reconstructed $\Delta x$ which matches the pre-defined values: 0 for $x<0$ and -10 pm for $x>0$. An example of a reconstruction when both $\Delta x$ and $\Delta z$ occur simultaneously with more details about the reconstruction procedure is provided in Supporting Information.

As one can see, it is easy to distinguish between the out-of-plane and in-plane atomic displacements even without performing a reconstruction by simply comparing the intensity contrast in the opposite CBED spots of monolayers or the regularity of interference fringes for bilayer samples. An out-of-plane defect will always result in a symmetric phase distribution between the mirror-symmetric CBED spots, and an in-plane defect – in an antisymmetric one.

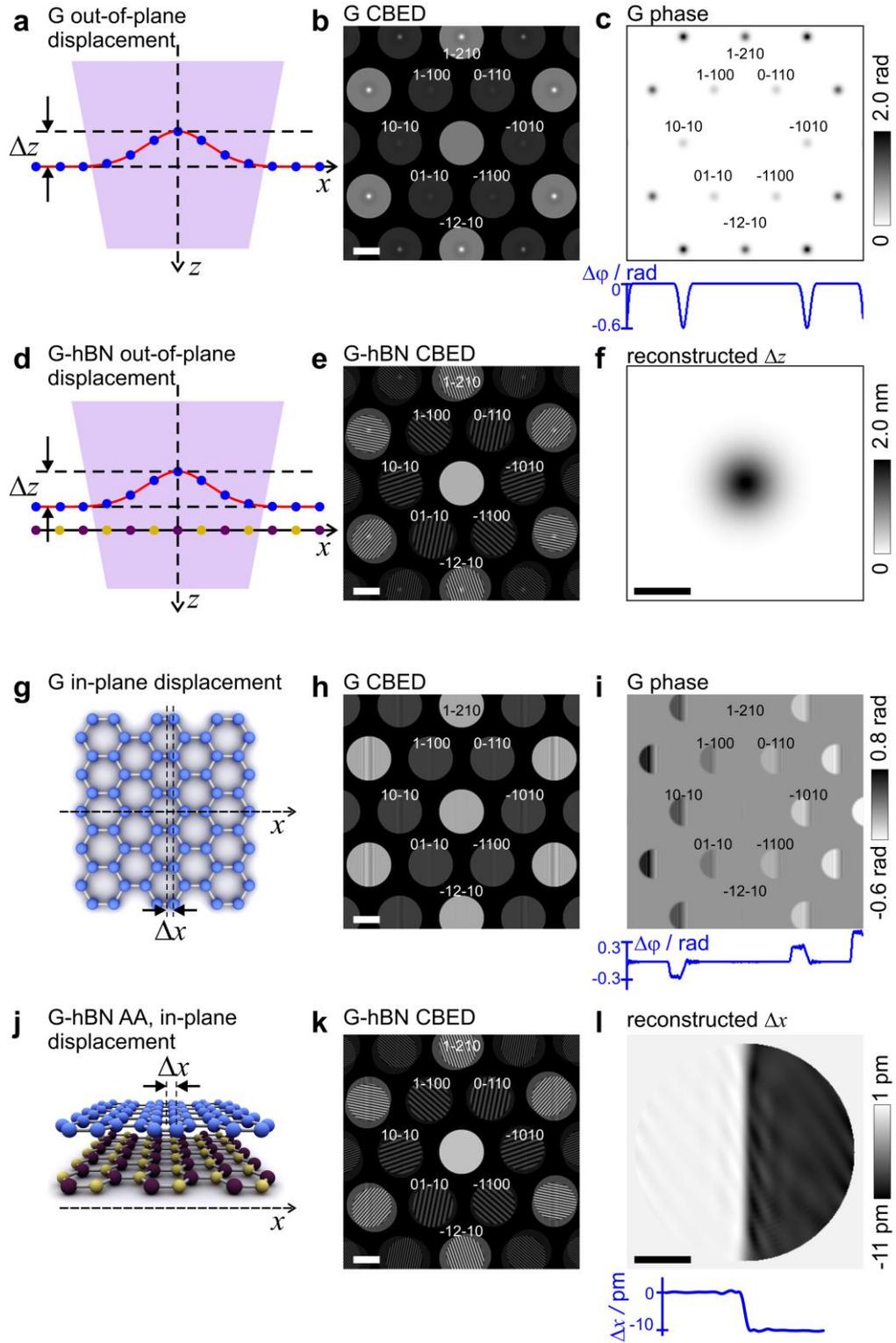

Fig. 3. Simulated CBED patterns for graphene and graphene-hBN bilayer heterostructures where graphene lattice is deformed.
(a) Side view illustration of a graphene layer with atoms displaced out of plane, not drawn to scale.
(b) Corresponding simulated CBED pattern where the atoms are displaced out of plane in the form of a bubble. The atomic $z$-positions are shifted by $\Delta z = -A_B \exp[-(x^2+y^2)/(2\sigma_B^2)]$, $A_B = 2$ nm, $\sigma_B = 2$ nm.

(c) The difference of the phases of the wavefronts scattered by graphene with and without the out-of-plane atomic displacement due to the bubble. The blue curve shows the profile through the center of the distribution that is at $K_y=0$.
(d) Sketch of the side view of a graphene-hBN bilayer with AA stacking area but with atoms in graphene displaced out of plane due to the presence of a bubble.
(e) Corresponding simulated CBED pattern for (d). The graphene atoms displacement same as in (b).
(f) Reconstructed distribution of the atomic out-of-plane displacement due to the bubble in the graphene layer. The scale bar corresponds to 5 nm.
(g) Sketch of the top view of the graphene layer with atoms displaced laterally (within the crystal plane).
(h) Simulated CBED pattern where the atoms positioned at $x>0$ are displaced by $\Delta x=-10$ pm.
(i) The difference between the phases of the wavefronts scattered from graphene with and without in-plane atomic displacement. The blue curve shows the profile through the centre of the distribution that is at $K_y=0$.
(j) Sketch of the side view of a graphene-hBN bilayer, AA stacking area, with atoms in graphene displaced within graphene plane as in (g).
(k) Corresponding simulated CBED pattern for graphene-hBN heterostructures described in (j).
(l) Reconstructed distribution of the atomic in-plane $\Delta x$ displacement in the graphene layer. The scale bar corresponds to 5 nm.
For these simulations the distance between the layers is 3.35 Å, $\Delta f=-2$ μm, and the imaged area is about 28 nm in diameter. The scale bars in (b), (e), (h) and (k) correspond to 2 nm$^{-1}$.

**Experimental results.** Fig. 4 shows CBED patterns for three of our samples: aligned graphene on hBN (Fig. 4a), graphene rotated with respect to hBN by a small angle (Fig. 4b), and a multilayered sample (Fig. 4c). As predicted by the simulations (Fig. 2), the interference patterns are tangential and radial in Fig. 4a and Fig. 4b respectively. The particular arrangement of the interference patterns (for instance, the number of interference fringes) depends on $\Delta f$ as well as the misorientation angle between the two crystals, among other parameters. The misorientation angle calculated for this particular sample is 2.5°±0.1° (see Supporting Information). More examples of experimental CBED patterns for graphene-hBN heterostructures are provided in Supporting Information.

Figure 4c shows a CBED pattern of a three layers system (hBN sandwiched between two layers of graphene) acquired at $\Delta f \approx 0$. Because $\Delta f$ is so small, the period of the interference fringes is very large and no interference fringes are observed within the overlapping CBED spots. From the position of the spots, we measure that the relative rotation between hBN and one of the graphene layers is 2.5°, and the relative rotation between the two graphene layers is 17°. Two of the three layers exhibit unperturbed crystalline structure as evident from the homogeneous intensity within their CBED spots. The intensity variations of opposite contrast in opposite CBED spots of the hBN layer indicate the presence of an in-plane strain (as indicated by the red arrows in Fig. 4c). The contrast variation on the circumference of each CBED discs in Fig. 4 and 5 are attributed to charging of the dust particles at the condenser aperture.

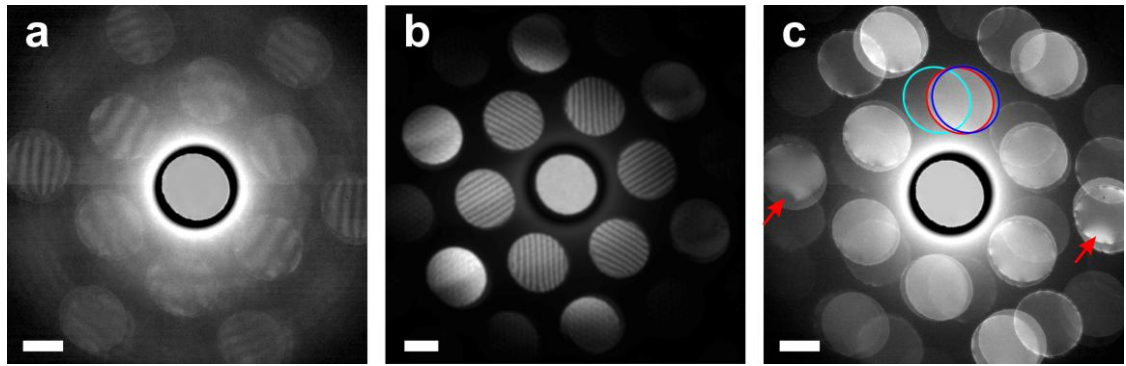

Fig. 4. Experimental CBED patterns of graphene-hBN samples. (a) CBED pattern where the direction of fringes indicates that there is almost no relative rotation between the layers, $\Delta f$=-5 μm, $\alpha$=6.93 mrad, which gives the diameter of the imaged area of about 70 nm. (b) CBED pattern of another sample where the direction of fringes indicate a relative rotation between the layers, $\beta$=2.5°, $\Delta f$=-3 μm, $\alpha$=8.023 mrad, which gives the diameter of the imaged area of about 48 nm. (c) CBED pattern of graphene/hBN/graphene heterostructure sample, $\Delta f \approx 0$, $\alpha$=8.023 mrad, the diameter of the imaged area is about 1 nm (more information on the distribution of the probing wavefront is provided in Supporting Information). The intensity of the central spot is reduced by factor 0.002 in (a), 0.1 in (b) and 0.005 in (c). The scale bars are 2 nm$^{-1}$.

In order to demonstrate the holographic nature of our CBED patterns, we imaged a ripple defect in stacking between the two layers. Such defects are often associated with basal plane dislocations and have also been referred to as ripplocations (23). Fig. 5a presents a CBED pattern from an area with a stacking fault between slightly misoriented graphene and hBN, which is evident by the presence of a distinctive ridge in the interference patterns in the first- and higher-order CBED spots (see also Fig. 5b). We note that no features are visible in the zero-order spot, suggesting that the defect induces only marginal additional absorption. However, this defect introduces a significant additional phase shift between the electron waves scattered from the top and bottom crystals, which is readily picked up in the CBED interference patterns. If it were not for this interference, quantitative imaging of such a defect would be next to impossible. CBED spots which originate from graphene are found at a slightly larger diffraction angle allowing differentiation of the spots corresponding to graphene from those for hBN. The overlap between CBED spots from the two crystals is less in the higher diffraction orders but the intensity contrast due to corrugation is more pronounced in the higher order CBED spots. Thus, by visual inspection of the high-order CBED spots, the type of corrugation and the layer with the corrugation can be readily identified. For example, in the CBED pattern shown in Fig. 5a, the ripple marked by the cyan arrows in Fig. 5a (also in magnified image in Fig. 5b) manifests itself identically in all the higher-order CBED spots, which suggests that this is an out-of plane ripple. Also, it is clearly seen in the third-order CBED spots, where the spots from hBN and graphene are sufficiently separated (as indicated by the yellow arrows in Fig. 5a), that the projection of the ripple exists only in one of the spots (hBN), unambiguously identifying the corrugated layer.

In order to extract quantitative information about the defect we performed holographic reconstruction of the CBED pattern image presented in Fig. 5a as described above. The phase unwrapping was applied by using a procedure introduced by Schofield and Zhu (24). Fig. 5c shows the recovered out-of-plane $\Delta z$ atomic displacements obtained from symmetric component of the CBED picture by averaging the reconstructed phase distributions from all six first-order CBED spots and applying Eq. 1. Fig. 5d shows the recovered in-plane $\Delta x$ atomic displacements obtained by extracting the antisymmetric component of the phase distribution from two opposite first-order CBED spots and by applying Eq. 3. Fig. 5e compares the out-of-plane and in-plane atomic shifts along the ripple. In our case the retrieved height of the out-of-plane ripple in hBN layer is about 2 nm. This is reasonable, since out of plane ripples are often observed for graphene/hBN stacks due to self-cleansing effects (25).

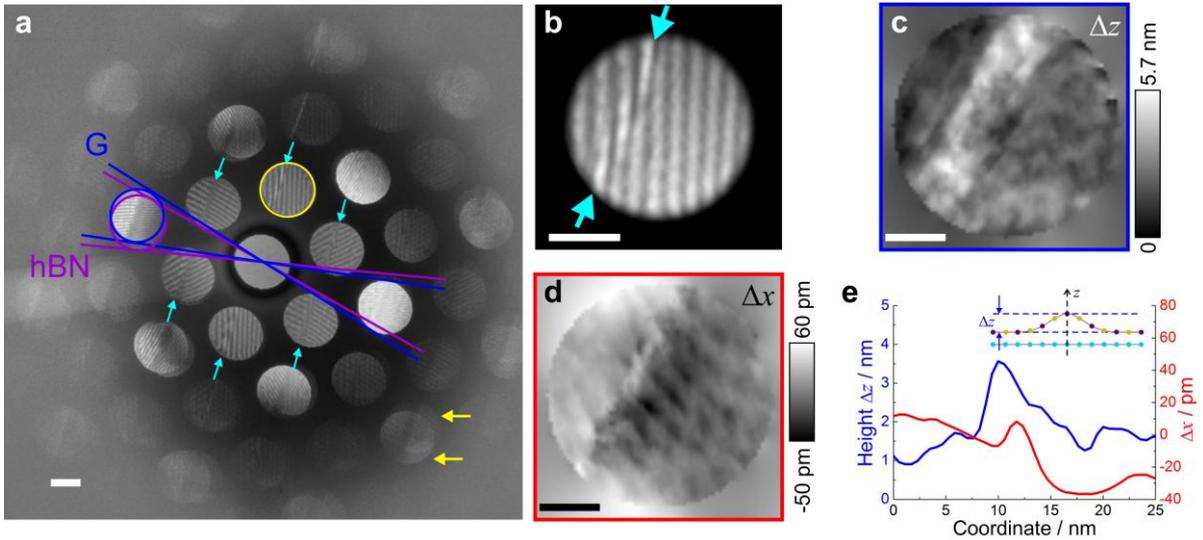

Fig. 5. Extracting the shape of the out-of-plane ripple from a CBED pattern. (a) Experimental CBED pattern acquired at $\Delta f$=-3 μm, with defects in the interference patterns marked by the arrows. The blue and purple lines indicate the relative rotation between graphene and hBN layers, which amounts to 3°. The cyan arrows indicate an out-of-plane ripple observed in the first-order CBED spots. The yellow arrows indicate the separation of CBED spots originating from graphene and hBN layers, where it becomes clear that the ripple is in the hBN layer. The intensity of the central spot is reduced by a factor of 0.1. The scale bar corresponds to 2 nm$^{-1}$. (b) Magnified selected spot (circled yellow in (a)) where irregularities of the fringe pattern can be seen. The scale bar corresponds to 1 nm$^{-1}$. (c) The reconstructed distribution of the ripple height $\Delta z$. (d) The reconstructed distribution of the lateral shift $\Delta x$. (e) Profiles for the magnitude of $\Delta z$ and $\Delta x$ profiles perpendicular to the ripple in (c) and (d). The scale bars in (c) and (d) correspond to 10 nm in real space.

## Discussion

We have demonstrated that single CBED patterns of van der Waals heterostructures allow for direct visualization of the three-dimensional atomic distribution in each individual layer. Even without reconstruction, qualitative information about the type (stretching or out-of-plane rippling) and the extent of lattice deformation can be directly obtained by simple comparison of intensity distributions in the opposite CBED spots. For bilayer materials, a holographic approach can be applied to quantitatively reconstruct the values of three-dimensional atomic displacements.

The resolution provided with our technique depends on the size of the imaged area, that is, on $\Delta f$ and the corresponding magnification. To evaluate the sensitivity of our method to atomic shifts we estimate that a phase shift of 0.1 rad is sufficient to cause detectable variations in intensity distribution. Such a phase shift can be caused by out-of-plane atomic shifts of $\Delta z \approx 0.35$ nm or by an in-plane atomic shift of $\Delta x \approx 3.4$ pm (in accordance to Eq. 2 and Eq. 3). Thus, the sensitivity to in-plane atomic displacements is about 100 times higher than the sensitivity to out-of-plane displacements. The sensitivity to spatial position is of the order of 1 nm when the location of atomic displacements are obtained by comparing intensity distributions from the opposite CBED spots and is a few nanometers when the holographic reconstruction is applied (as discussed in Supporting Information).

Quantitative information in the $z$ direction (parallel to the electron beam) is notoriously difficult to obtain from a projection imaging technique such as transmission electron microscopy. Thus the ability to obtain quantitative information about the relative location of the atomic sheets in the z-direction is one of the strengths of our approach. Furthermore, our approach does not require any special sample preparation – it can be applied to any samples being observed in TEM in traditional modes. Classical electron tomography methods used to gain three-dimensional positional data struggle except when the target is a perfectly crystalline nano-object that can be imaged along several zone axes (26). Van Dyck et al have demonstrated that atomic resolution three-dimensional coordinates could be achieved from only a single projection using a combination of exit wave reconstruction from a focal series of images and a "Big Bang" analysis of the quantitative phase shift for individual atomic columns (27). Their approach requires a sequence of atomically resolved high-resolution images (to recover the complex-valued exit wave which is a complicated analysis by itself) and only obtains a single z-location for each atom column position. By contrast our approach gains high resolution z-positional data on the relative position of the two separate layers from a *single* CBED pattern. Although currently the lateral resolution we obtain is poorer than that demonstrated by Van Dyck et al, our approach requires orders of magnitude lower electron exposures so it has a potential to be applied to heterostructures containing beam-sensitive 2D materials and even overlapping protein membranes (28). In addition, our approach can be further advanced by obtaining many CBED patterns to reconstruct a large area (diffractive imaging). Also, depending on $\Delta f$, the area studied by our approach can be tuned from gaining sensitive relative atomic displacements for just a few nm$^2$, to regions ~1 μm wide, as is typical of the active area of lithographically patterned 2D heterostructure devices.

Our results demonstrate that CBED on van der Waals heterostructures can be applied to yield a plethora of information about the stacks. This technique has not previously been applied to van der Waals heterostuctures but we have shown that it is highly versatile and as it can be performed on any conventional TEM instrumentation. We expect that this approach will find a progressive use in the expanding field of 2D materials. Furthermore it could be extended to the analysis of dopant atoms, local oxidation or trapped species within the heterostructures, as all of these will affect the interference patterns obtained.

## Materials and Methods

The samples were obtained by mechanical stacking of mechanically exfoliated graphene and hBN layers on a Si/SiO$_2$ substrate. By using "pick and lift" technique (29) - the whole stack was then transferred on a quantifoil carbon support film for observation in the TEM. CBED was realized in a probe side aberration corrected scanning TEM at an accellerating voltage of 80 keV and a convergence semi angle, $\alpha$, of ~6 – 8 mrad. During experiment, the focal lengths of the objective and condenser lenses were kept constant, thus there was no change in the convergence angle. The sample *z*-position was changed by moving the sample along the optical axis. Bilayer structure significantly improves the stability of 2D crystals upon exposure to high-energy electrons and the electron dose required for CBED imaging is low, so no evidence of knock-on damage was observed during prolonged data acquisition (30, 31).

**Acknowledgements**

This work was supported by the European Union Graphene Flagship Program, European Research Council Synergy Grant "Hetero2D" and European Research Council Starting Grant "EvoluTEM"; the Royal Society; Engineering and Physical Research Council (UK); and US Army Research Office (grant W911NF-16-1-0279). S.J.H and E.P. acknowledge funding from the Defense Threat Reduction Agency (grant HDTRA1-12-1-0013) and the Engineering and Physical Sciences Research Council (UK) (grants EP/K016946/1, EP/L01548X/1, EP/M010619/1 and EP/P009050/1).

# References

1. Geim AK & Grigorieva IV (2013) Van der Waals heterostructures. *Nature* 499(7459):419-425.
2. Novoselov KS, Mishchenko A, Carvalho A, & Castro Neto AH (2016) 2D materials and van der Waals heterostructures. *Science* 353(6298):aac9439-aac9439.
3. Haigh SJ*, et al.* (2012) Cross-sectional imaging of individual layers and buried interfaces of graphene-based heterostructures and superlattices. *Nat. Mater.* 11(9):764–767.
4. Argentero G*, et al.* (2017) Unraveling the 3D Atomic Structure of a Suspended Graphene/hBN van der Waals Heterostructure. *Nano Lett.* 17(3):1409-1416.
5. Kossel W & Möllenstedt G (1939) Elektroneninterferenzen im konvergenten Bündel. *Ann. Phys.* 36:113.
6. Goodman P (1975) A practical method of three-dimensional space-group analysis using convergent-beam electron diffraction. *Acta Crystallogr. Sect. A* 31(6):804–810.
7. Buxton BF, Eades JA, Steeds JW, & Rackham GM (1976) The symmetry of electron diffraction zone axis patterns. *Philos. Trans. Royal Soc. A* 281(1301):171–194.
8. Tanaka M, Sekii H, & Nagasawa T (1983) Space-group determination by dynamic extinction in convergent-beam electron diffraction *Acta Crystallogr. Sect. A* 39(NOV):825–837.
9. Tanaka M, Saito R, & Sekii H (1983) Point-group determination by convergent-beam electron diffraction. *Acta Crystallogr. Sect. A* 39(MAY):357–368.
10. Jones PM, Rackham GM, & Steeds JW (1977) Higher-order Laue zone effects in electron-diffraction and their use in lattice-parameter determination. *Proc. R. Soc. London Ser. A-Math. Phys. Eng. Sci.* 354(1677):197-222.
11. Carpenter RW & Spence JCH (1982) Three-dimensional strain-field information in convergent beam electron diffraction patterns. *Acta Crystallogr. Sect. A* 38(JAN):55–61.
12. Kelly PM, Jostsons A, Blake RG, & Napier JG (1975) The determination of foil thickness by scanning transmission electron microscopy. *physica status solidi (a)* 31(2):771–780.
13. Champness PE (1987) Convergent beam electron diffraction. *Mineral. Mag.* 51(359):33–48.
14. Tanaka M, Saito R, Ueno K, & Harada Y (1980) Large-angle convergent-beam electron diffraction. *J. Electron Microsc.* 29(4):408–412.
15. Morniroli JP (2006) CBED and LACBED characterization of crystal defects. *J. Microsc.-Oxf.* 223:240–245.
16. Gabor D (1948) A new microscopic principle. *Nature* 161(4098):777–778.
17. Gabor D (1949) Microscopy by reconstructed wave-fronts. *Proc. R. Soc. A* 197(1051):454–487.
18. Meyer JC*, et al.* (2007) On the roughness of single- and bi-layer graphene membranes. *Solid State Communications* 143(1-2):101-109.
19. Leith EN & Upatnieks J (1962) Reconstructed wavefronts and communication theory. *J. Opt. Soc. Am.* 52(10):1123–1130.
20. Leith EN & Upatnieks J (1963) Wavefront reconstruction with continuous-tone objects. *J. Opt. Soc. Am.* 53(12):1377–1381.
21. Mollenstedt G & Wahl H (1968) Electron holography and reconstruction with laser light. *Naturwissenschaften* 55(7):340–341.
22. Lehmann M & Lichte H (2002) Tutorial on off-axis electron holography. *Microsc. Microanal.* 8(6):447–466.
23. Kushima A, Qian X, Zhao P, Zhang S, & Li J (2015) Ripplocations in van der Waals Layers. *Nano Lett.* 15(2):1302-1308.
24. Schofield MA & Zhu YM (2003) Fast phase unwrapping algorithm for interferometric applications. *Opt. Lett.* 28(14):1194–1196.
25. Kretinin AV*, et al.* (2014) Electronic properties of graphene encapsulated with different two-dimensional atomic crystals. *Nano Lett.* 14(6):3270-3276.
26. Van Aert S, Batenburg KJ, Rossell MD, Erni R, & Van Tendeloo G (2011) Three-dimensional atomic imaging of crystalline nanoparticles. *Nature* 470:374–377.

27. Van Dyck D & Chen F-R (2012) 'Big Bang' tomography as a new route to atomic-resolution electron tomography. *Nature* 486(7402):243–246.
28. Nair RR*, et al.* (2010) Graphene as a transparent conductive support for studying biological molecules by transmission electron microscopy. *Appl. Phys. Lett.* 97(15):153102
29. Wang L*, et al.* (2013) One-dimensional electrical contact to a two-dimensional material. *Science* 342(6158):614-617.
30. Zan R*, et al.* (2013) Control of radiation damage in MoS2 by graphene encapsulation. *ACS Nano* 7(11):10167-10174.
31. Algara-Sillera G, Kurascha S, Sedighi M, Lehtinen O, & Kaiser U (2013) The pristine atomic structure of MoS2 monolayer protected from electron radiation damage by graphene. *Appl. Phys. Lett.* 103(20):203107.

Supporting Information for

# Convergent beam electron holography for analysis of van der Waals heterostructures

*Tatiana Latychevskaia[1], Colin Robert Woods[2,3], Yi Bo Wang[2,3], Matthew Holwill[2,3], Eric Prestat[4], Sarah J. Haigh[2,4], Kostya S. Novoselov[2,3]*

[1]*Institute of Physics, Laboratory for ultrafast microscopy and electron scattering (LUMES), École Polytechnique Fédérale de Lausanne (EPFL) , CH-1015 Lausanne, Switzerland*

[2]*National Graphene Institute, University of Manchester, Oxford Road, Manchester, M13 9PL, UK*

[3]*School of Physics and Astronomy, University of Manchester, Oxford Road, Manchester, M13 9PL, UK*

[4]*School of Materials, University of Manchester, Oxford Road, Manchester, M13 9PL, UK*

# Contents

# 1. Simulation procedure

## 1.1 Transmission function of monolayer

The transmission function of an atomic lattice for approximation of a weak phase object is given by

$$T(\vec{r}) = \exp[i\sigma V_z(\vec{r})] \approx 1 + i\sigma v_z(\vec{r}) \otimes L(\vec{r}), \qquad (S1)$$

where $V_z(\vec{r})$ is the projected potential of the entire sample, $v_z(\vec{r})$ is the projected potential of an individual atom, $L(\vec{r})$ is the function describing positions of the atoms in lattice, $\vec{r} = (x, y, z)$ is the coordinate in the layer, and $\otimes$ denotes convolution. The interaction parameter $\sigma$ is given by

$$\sigma = \frac{2\pi m e \lambda}{h^2} = \frac{2\pi e}{h \upsilon},$$

where $m$ is the relativistic mass of electron, $e$ is an elementary charge, $\lambda$ is wavelength of the electrons, $\upsilon$ means the electron velocity, and $h$ is Planck's constant.

## 1.2 CBED on monolayer: Far-field wavefront distribution

A spherical wavefront $\psi_0(\vec{r}) = \dfrac{\exp(\pm ikr)}{r}$, where $k = \dfrac{2\pi}{\lambda}$, and "$-$" corresponds to a convergent wave while "$+$" corresponds to a divergent wave, incidents the layer described by the transmission function $T(\vec{r})$. The exit wave immediately after the specimen is given by

$$\psi(\vec{r}) = \psi_0(\vec{r}) T(\vec{r}).$$

The distribution of the scattered wave in the far-field $\Psi(\vec{R})$ is given by ($r \ll R$)

$$\Psi(\vec{R}) \propto \frac{\exp(ikR)}{R} \int \psi(\vec{r}) \exp\left(-ik \frac{\vec{r}\vec{R}}{R}\right) d\vec{r},$$

where $\vec{R} = (X, Y, Z)$ is the coordinate in the detector plane, and where approximation $|\vec{r} - \vec{R}| \approx R - \dfrac{\vec{r}\vec{R}}{R}$ was applied. Thus, the wavefront distribution in the far-field is given by the Fourier transform (FT) of the exit wave which can be re-written as:

$$\Psi \propto \mathrm{FT}(\psi) = \mathrm{FT}(\psi_0 T) = \mathrm{FT}(\psi_0) + i\sigma \mathrm{FT}\{\psi_0 \cdot [v_z(\vec{r}) \otimes L(\vec{r})]\} =$$
$$= \mathrm{FT}(\psi_0) + i\sigma \mathrm{FT}(\psi_0) \otimes [\mathrm{FT}(L) \mathrm{FT}(v_z)]$$

where we used Eq. S1. When the incident wavefront is a plane wave, $\psi_0(\vec{r}) = 1$, the far-field distribution of the scattered wave is described by $\mathrm{FT}(L)\mathrm{FT}(v_z)$. The first term, $\mathrm{FT}(L)$, is the reciprocal lattice or the distribution of the diffraction peaks. The second term, $\mathrm{FT}(v_z)$, is related to the far-field distribution of the wavefront scattered on an individual atom, it is described by a slow varying function when compared to the first term, and therefore provides a slow varying modulation for the entire diffraction pattern.

When the incident wavefront is a convergent/divergent wavefront, each diffraction peak turns into a finite size convergent beam electron diffraction (CBED) spot that is a holographic image of the specimen. CBED pattern is then described by convolution of $\text{FT}(L)$ with $\text{FT}(\psi_0)$. The intensity variations within a selected CBED spot are described by $\text{FT}(L)$, which is solely defined by the arrangement of atoms $L(\vec{r})$. Therefore, in the simulation we only vary the atomic positions as described by $L(\vec{r})$.

## 1.3 Applied simulation routine

To study CBED spot intensity distribution as a function of atomic displacements we create distribution of atoms with pre-defined positions $L(\vec{r})$ and we simulate the far-field distribution of the scattered wavefront as:

$$\Psi_l(K_x, K_y) = \sum_i L(\vec{r}_i)\psi_0(\vec{r}_i)\exp\left[-i(K_x x_i + K_y y_i)\right]\exp\left(-i z_i \sqrt{K^2 - K_x^2 - K_y^2}\right), \quad (S2)$$

where $K$-coordinates are introduced as $\vec{K} = (K_x, K_y, K_z) = k\dfrac{\vec{R}}{R} = \dfrac{2\pi}{\lambda R}(X, Y, Z)$, $|\vec{K}| = k = \dfrac{2\pi}{\lambda}$, $K_z = \sqrt{K^2 - K_x^2 - K_y^2}$, $i$ runs through all the atoms in the lattice, $\vec{r}_i$ are the coordinates of the atoms, $l$ is the lattice number, and $\psi_0(\vec{r})$ is the incident wavefront distribution. $\psi_0(\vec{r})$ is calculated by simulation diffraction of the spherical wavefront on a limiting aperture (second condenser aperture) positioned at a plane $\vec{r}_0$:

$$\psi_0(\vec{r}) = \iint a(\vec{r}_0)\dfrac{\exp(-ikr_0)}{r_0}\dfrac{\exp(ik|\vec{r}_0 - \vec{r}|)}{|\vec{r}_0 - \vec{r}|}d\vec{r}_0,$$

where $a(\vec{r}_0)$ is the aperture function. For further details about the incident wavefront distributions see section "Probing wave distribution".

In the simulations no fast Fourier transforms are applied. The simulations are done by calculating divergent incident wavefront scattered off individual atoms in lattices (Eq. S2) and summing up the scattered complex-valued waves in the detector plane. For two or more lattices, the calculated complex-valued far-field wavefront distributions are added together and the squared absolute value of the result gives the total intensity distributions.

### 1.4 Probing wave distribution

Figure S1 shows the probing electron wave distribution at different planes. At some plane the wavefront passes through a limiting aperture (second condenser aperture in our experiment) whose image is formed before the sample plane, as sketched in Fig. S1a. Figure S1b shows intensity distribution of a zero-order CBED spot at the detector plane. Because the electron beam in a TEM is partially coherent, a zero-order CBED spot is in fact an inline hologram. The edges of the zero-order diffraction spot exhibit interference patterns similar to that resulting from diffraction on an edge, see Fig. S1c. Such an inline hologram can be reconstructed by propagating the wavefront backwards. A sharp in-focus reconstruction of the edge of the aperture is found at a distance of $\Delta f = -45$ μm (Fig. S1d–e), the other fringes are due to the twin image (1). Thus, the probing wave can be represented as a convergent spherical wave which is limited by a round aperture at $\Delta f = -45$ μm. The presence of the aperture leads to the corresponding modulation of the probing wave at other planes. For example, the intensity distribution of the probing wave in the sample plane at $\Delta f = -2$ μm exhibits concentric interference fringes, Fig. S1f. To mimic the experimental conditions in the simulations, the probing wave was modelled as a convergent spherical wavefront passing through an aperture of 640 nm in diameter at $\Delta f = -45$ μm. The intensity distribution of the simulated probing wave at the sample plane at a typical sample *z*- position $\Delta f = -2$ μm is shown in Fig. S1g. The intensity distribution of the simulated probing wave at the sample plane when $\Delta f = 0$ is shown in Fig. S1h, demonstrating that a finite area of the sample of about 1 nm in diameter can be probed.

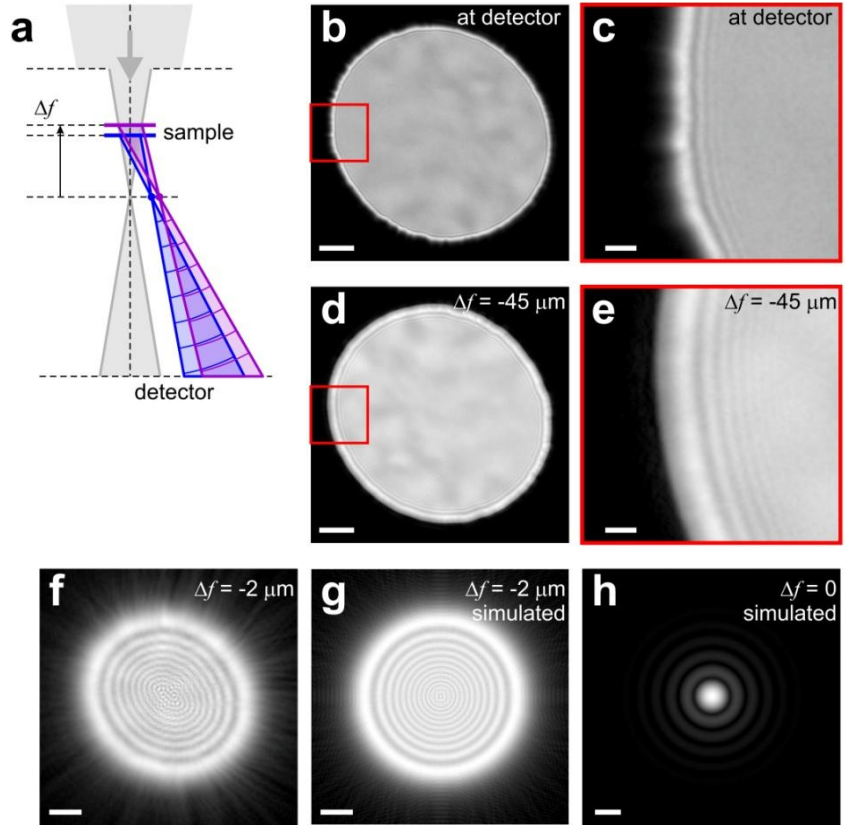

Fig. S1. Experimental and simulated images of the probing wave distribution at different planes in the transmission electron microscope.
(a) Schematic representation of electron wave propagation in CBED mode.
(b) Experimentally measured intensity distribution in the zero-order CBED spot where the *z*-position of the sample is $\Delta f = -2$ μm, and its magnified region of the edge (c). (d) Amplitude of the wavefront spot at $\Delta f = -45$ μm and its magnified region (e), obtained by backward propagation of the wavefront from the detector plane. (f) Amplitude of the wavefront spot at $\Delta f = -2$ μm, obtained by backward propagation of the wavefront from the detector plane. (g) Amplitude of the simulated wavefront at $\Delta f = -2$ μm. (h) Amplitude of the simulated wavefront at $\Delta f = 0$. The scalebars in (b) and (d) correspond to 100 nm, (c) and (e) to 20nm, in (f) and (g) to 5 nm, and in (h) to 0.5 nm.

## 2. Positions of virtual sources

We consider a lattice illuminated by a convergent/divergent wavefront. The lattice is positioned at a distance $\Delta f$ from the plane where the incident wavefront converges to a point (virtual source plane). The lattice can be described by a two-dimensional Dirac comb function $L(\vec{r})$, where $\vec{r} = (x, y, \Delta f)$ is the coordinate in the lattice plane, and $\vec{a}_1$ and $\vec{a}_2$ are the translation vectors. A convergent/divergent wavefront is described by $\exp(i\alpha kr)$, where $\alpha = -1$ for a convergent wavefront which corresponds to underfocus CBED ($\Delta f < 0$), and $\alpha = +1$ for a divergent wavefront which corresponds to overfocus CBED ($\Delta f > 0$). The exit wave immediately after the lattice is given by $U(\vec{r}) = \exp(i\alpha kr) L(\vec{r})$. The wavefront propagated to the virtual source plane $(u, w, 0)$ can be calculated by the Huygens-Fresnel integral transform:

$$U_0(\vec{\rho}) = -\frac{i}{\lambda} \iint \exp(i\alpha kr) L(\vec{r}) \frac{\exp(-i\alpha k |\vec{r} - \vec{\rho}|)}{|\vec{r} - \vec{\rho}|} \, d\vec{r}, \tag{S3}$$

where $\vec{\rho}$ is the coordinate in the virtual source plane and where the following approximation can be applied

$$|\vec{r} - \vec{\rho}| \approx r - \frac{\vec{r}\vec{\rho}}{r} + \frac{\rho^2}{2r}$$

and Eq. S3 can be re-written as

$$U_0(\vec{\rho}) \approx -\frac{i}{\lambda} \iint L(\vec{r}) \frac{1}{r} \exp\left(i\alpha k \frac{\vec{r}\vec{\rho}}{r}\right) \exp\left(-i\alpha k \frac{\rho^2}{2r}\right) d\vec{r} =$$

$$= \frac{i}{\lambda \Delta f} \exp\left(-ik \frac{\rho^2}{2\Delta f}\right) \iint L(\vec{r}) \exp\left(ik \frac{\vec{r}\vec{\rho}}{\Delta f}\right) d\vec{r}. \tag{S4}$$

The integral in Eq. S4 is a Fourier transform of the lattice function $L(\vec{r})$ and the result is the reciprocal lattice defined as:

$$\tilde{L}(\vec{k}) = \iint L(\vec{r}) \exp(i\vec{k}\vec{r}) \, d\vec{r}.$$

Thus we obtain

$$U_0(\vec{\rho}) \approx \frac{i}{\lambda \Delta f} \exp\left(-\frac{i\pi}{\lambda \Delta f} \rho^2\right) \tilde{L}\left(\frac{\vec{\rho}}{\lambda \Delta f}\right). \tag{S5}$$

Eq. S5 describes the distribution of the virtual sources in the virtual source plane. Each virtual source has additional phase factor $\exp\left(-\frac{i\pi}{\lambda \Delta f} \rho^2\right)$ that corresponds to an incident convergent/divergent wavefront. Without this phase factor, the far-field diffracted wave would be given by a Dirac comb function describing position of individual diffraction peaks. With this phase factor, each diffraction peak is turned into a finite-sized CBED spot.

For diffraction from a hexagonal lattice, the six virtual sources are positioned in the $(u, w, 0)$-plane with six-fold symmetry at a distance

$$\rho^{(i)} = |\Delta f| \tan \vartheta^{(i)} \approx \frac{\lambda |\Delta f|}{a^{(i)}}$$

from $(0,0,0)$-coordinate. Here $\vartheta^{(i)}$ is the diffraction angle corresponding to lattice period $a^{(i)}$ of lattice $i$. For example, for $\Delta f = 2$ μm, $\rho^{(G)} = 39.2$ nm and $\rho^{(BN)} = 38.4$ nm.

## 3. Interference distribution in a CBED spot of bilayer

When imaging a bilayer, the interference pattern within each CBED spot is analogous to an interference pattern created by two waves originating from two virtual sources, as illustrated in Fig. 1 (main text). We set symbol (1) for lattice1 and symbol (2) for lattice2. We assume that lattice1 is shifted by $\left(\Delta x^{(1)}, \Delta y^{(1)}\right)$ and lattice2 is shifted by $\left(\Delta x^{(2)}, \Delta y^{(2)}\right)$ relatively to a centred lattice (a lattice in which one of its hexagons is centred at the origin of the $(x, y)$-plane). We also assume that the lattices are separated by a distance $d$. Lattice1 is positioned at a distance $\Delta f$ from the virtual source plane, lattice2 is positioned at a distance $(\Delta f + d)$ from the virtual source plane, Fig. 1a (main text). The two corresponding wavefront distributions in the $(u, w)$-plane are given by Eq. S5:

$$U^{(1)}(\vec{\rho}) \propto \frac{1}{\Delta f} \exp\left(-\frac{\pi i}{\lambda \Delta f}\left(\rho^{(1)}\right)^2\right) \delta\left(\vec{\rho} - \vec{\rho}_m^{(1)}\right),$$

$$U^{(2)}(\vec{\rho}) \propto \frac{1}{(\Delta f + d)} \exp\left(-\frac{\pi i}{\lambda(\Delta f + d)}\left(\rho^{(2)}\right)^2\right) \delta\left(\vec{\rho} - \vec{\rho}_m^{(2)}\right).$$

where $\vec{\rho}_m^{(1)}$ and $\vec{\rho}_m^{(2)}$ are the position of the virtual sources in the $(u, w)$-plane, and $m$ ($m = 1...6$) is a CBED spot number.

Next we consider propagation of the wavefronts from the $(u, w)$-plane to the detector $(X, Y)$-plane. Each virtual source creates a divergent spherical wave described by $\exp\left(ik\left|\vec{R} - \vec{\rho}_m\right|\right)$. The interference pattern within a CBED is described as

$$I_m(R) \propto$$

$$\propto \left|\exp\left(-\frac{i\pi}{\lambda \Delta f}\left(\rho^{(1)}\right)^2\right)\exp\left(ik\left|\vec{R} - \vec{\rho}_m^{(1)}\right| + i\gamma_m^{(1)}\right) + \exp\left(-\frac{i\pi}{\lambda(\Delta f + d)}\left(\rho^{(2)}\right)^2\right)\exp\left(ik\left|\vec{R} - \vec{\rho}_m^{(2)}\right| + i\gamma_m^{(2)}\right)\right|^2, \quad (S6)$$

where $\gamma_m^{(1)}$ and $\gamma_m^{(2)}$ are the constant phases of the virtual sources defined as following. When the lattice is shifted by $\left(\Delta x^{(i)}, \Delta y^{(i)}\right)$, its inverse Fourier transform gains an addition linear phase shift:

$$\exp\left[\frac{2\pi i}{\lambda \Delta f}\left(u\Delta x^{(i)} + w\Delta y^{(i)}\right)\right].$$

This phase shift is additional to the complex-valued spectrum of a centred lattice. For example, when lattice $i$ is shifted by $\left(\Delta x^{(i)}, \Delta y^{(i)}\right)$, the corresponding virtual source at $\vec{\rho}_m^{(i)} = \left(u_m^{(i)}, w_m^{(i)}\right)$ gains a constant phase shift

$$\exp\left[\frac{2\pi i}{\lambda \Delta f}\left(u_m^{(i)} \Delta x^{(i)} + w_m^{(i)} \Delta y^{(i)}\right)\right] = \exp\left[i\gamma_m^{(i)}\right],$$

and thus

$$\gamma_m^{(i)} = \frac{2\pi}{\lambda \Delta f}\left(u_m^{(i)} \Delta x^{(i)} + w_m^{(i)} \Delta y^{(i)}\right).$$

In Eq. S6 the following approximation can be applied:

$$\left|\vec{R} - \vec{\rho}\right| = \sqrt{R^2 - 2\vec{R}\vec{\rho} + \rho^2} \approx R - \vec{R}\vec{\rho}/R + \rho^2/2R.$$

and Eq. S6 can be re-written as:

$$I_m(R) \propto$$

$$\propto \left|\exp\left[-\frac{ik}{2\Delta f}\left(\rho^{(1)}\right)^2\right]\exp\left(ik\left|\vec{R} - \vec{\rho}_m^{(1)}\right| + i\gamma_m^{(1)}\right) + \exp\left[-\frac{ik}{(\Delta f + d)}\left(\rho^{(2)}\right)^2\right]\exp\left(ik\left|\vec{R} - \vec{\rho}_m^{(2)}\right| + i\gamma_m^{(2)}\right)\right|^2 =$$

$$= \left|\exp\left[-ik\frac{\vec{R}\vec{\rho}_m^{(1)}}{R} + ik\frac{\left(\rho^{(1)}\right)^2}{2R} - ik\frac{\left(\rho^{(1)}\right)^2}{2\Delta f} + i\gamma_m^{(1)}\right] + \exp\left[-ik\frac{\vec{R}\vec{\rho}_m^{(2)}}{R} + ik\frac{\left(\rho^{(2)}\right)^2}{2R} - ik\frac{\left(\rho^{(2)}\right)^2}{2(\Delta f + d)} + i\gamma_m^{(2)}\right]\right|^2 =$$

$$= \left|\exp\left[ik\frac{\vec{R}\vec{\rho}_m^{(1)}}{R} - ik\frac{\left(\rho^{(1)}\right)^2}{2R} + ik\frac{\left(\rho^{(1)}\right)^2}{2\Delta f} - i\gamma_m^{(1)}\right] + \exp\left[ik\frac{\vec{R}\vec{\rho}_m^{(2)}}{R} - ik\frac{\left(\rho^{(2)}\right)^2}{2R} + ik\frac{\left(\rho^{(2)}\right)^2}{2(\Delta f + d)} - i\gamma_m^{(2)}\right]\right|^2 =$$

$$= 2 + 2\cos\left\{k\frac{\vec{R}}{R}\left(\vec{\rho}_m^{(1)} - \vec{\rho}_m^{(2)}\right) - \frac{k}{2R}\left[\left(\rho^{(1)}\right)^2 - \left(\rho^{(2)}\right)^2\right] + \frac{k}{2}\left[\frac{\left(\rho^{(1)}\right)^2}{\Delta f} - \frac{\left(\rho^{(2)}\right)^2}{\Delta f + d}\right] - \Delta\gamma_m\right\} \approx$$

$$2 + 2\cos\left\{\vec{K}\left(\vec{\rho}_m^{(1)} - \vec{\rho}_m^{(2)}\right) + \frac{\pi}{\lambda}\left[\frac{\left(\rho^{(1)}\right)^2}{\Delta f} - \frac{\left(\rho^{(2)}\right)^2}{\Delta f + d}\right] - \Delta\gamma_m\right\}, \tag{S7}$$

where we introduced

$$\Delta\gamma_m = \gamma_m^{(1)} - \gamma_m^{(2)}$$

and $\vec{K} = k\frac{\vec{R}}{R}$. Equation S7 is a general formula which describes the interference pattern in a CBED spot of a bilayer. In Eq. S7 the first term in the argument of cosine describes the interference fringes. The second term in the argument of cosine in Eq. S7 is a constant offset which however is not small and cannot be neglected. $\Delta\gamma_m$ depends on the local stacking of the layers (for example, AA or AB stacking) and defines the position of the centre of the interference pattern. If the local stacking under the centre of the electron beam is AA, then $\Delta\gamma_m = 0$ and the interference pattern has maxima at the centre of the CBED spots.

# 4 Period and tilt of fringes in a CBED spot of bilayer

## 4.1 Period of fringes in a CBED spot

The period of fringes in a CBED spot is given by the distance between the virtual sources (as also evident from the argument of cosine in Eq. S7). For two layers with relative rotation $\beta$ (Fig. S2a), the distance between the virtual sources is given by (Fig. S2b):

$$\Delta\rho^2 = \rho_1^2 + \rho_2^2 - 2\rho_1\rho_2 \cos\beta, \tag{S8}$$

where $\rho_1 = \Delta f \tan \vartheta^{(1)}$, $\rho_2 = (\Delta f + d) \tan \vartheta^{(2)}$, and $\vartheta^{(1)}$ and $\vartheta^{(2)}$ are the diffraction angles corresponding to the lattice periods in layers 1 and 2. The wavefront propagated from the virtual source $\vec{\rho}_i$ to the detector plane is given by

$$\exp\left(ik\left|\vec{R}-\vec{\rho}_i\right|\right) \approx \exp(ikR)\exp\left(-ik\vec{R}\vec{\rho}_i / R\right)$$

and the intensity of the interference pattern created by superposition of the two waves is described by

$$I(\vec{K}) \sim 2\cos\left[\vec{K}(\vec{\rho}_1 - \vec{\rho}_2)\right],$$

where $\vec{K} = k\vec{R}/R$.

Next, we consider an interference pattern created by two sources separated by $\Delta\rho$. For simplicity, we assume that the two sources are positioned at $\left(-\dfrac{\Delta\rho}{2},0\right)$ and $\left(\dfrac{\Delta\rho}{2},0\right)$. The intensity distribution of the interference pattern becomes:

$$I(\vec{K}) \sim 2\cos\left[\vec{K}(\vec{\rho}_1 - \vec{\rho}_2)\right] = 2\cos(\Delta\rho K_x),$$

and the period of the interference pattern is given by

$$T = \frac{2\pi}{\Delta\rho}, \tag{S9}$$

where $\Delta\rho$ is given by Eq. S8.

At small distances between the layers (a few Ångstrom), the period of fringes is very large so that no interference pattern is observed in a CBED spot. At large distances between the layers, the period gets smaller, and an interference pattern appears in CBED spot, as confirmed by the simulations shown in Fig. S2c–d.

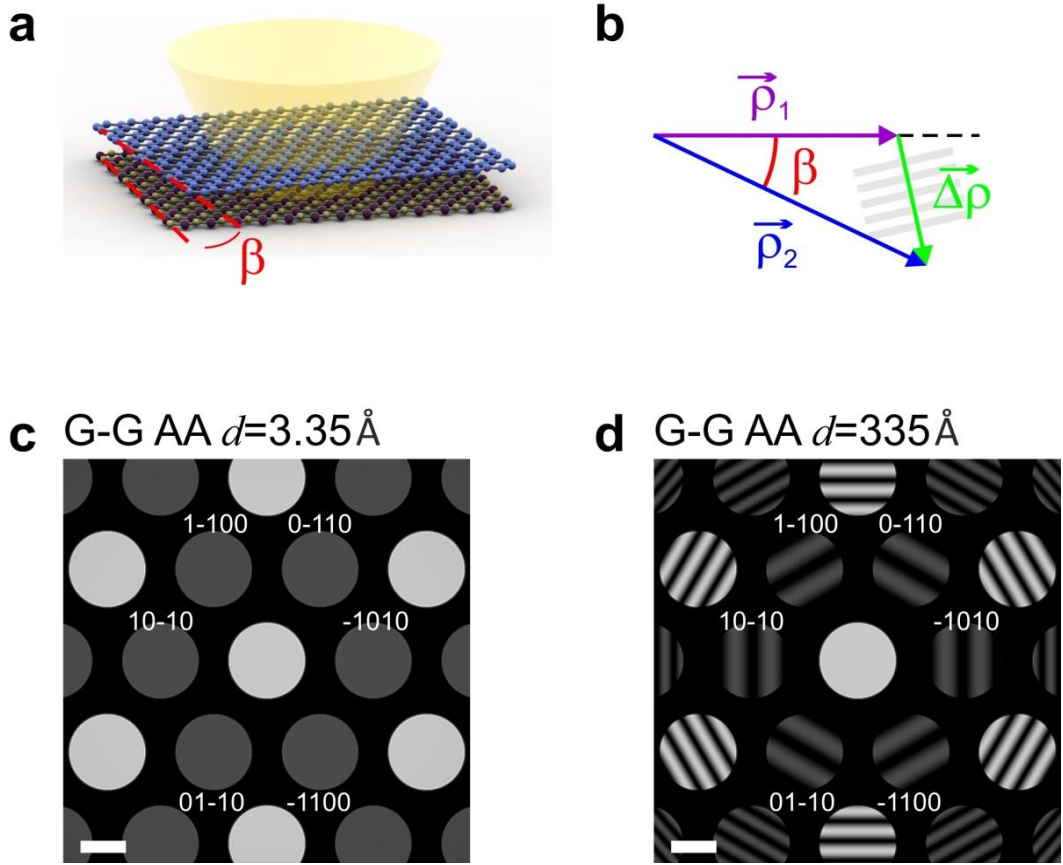

Fig. S2. Simulated CBED patterns of bilayer samples at different distance between the layers. (a) Schematic arrangement of 2D crystals and the incident electron beam. (b) Arrangement of the vectors in the virtual source plane. (c) Simulated CBED patterns of bilayer graphene, the distance between the layers is (c) 3.35 Å and (d) 335 Å. For these simulations the imaged area is about 50 nm in diameter, $\Delta f = -3$ μm, the number of pixels is 512 × 512 and $\Delta K = 3.51 \times 10^7$ m$^{-1}$. The scale bars in (c) and (d) correspond to 2 nm$^{-1}$.
11

## 4.2 Tilt of interference fringes in CBED spot

We assume that relative rotation between two layers is given by angle $\beta$ as illustrated in Fig. S3a. The tilt of the interference fringes, $\mu$, is readily found from the geometrical arrangement of the vectors in the virtual source plane (Fig. S3b):

$$\tan\mu = \frac{\rho_2 \sin\beta}{\rho_2 \cos\beta - \rho_1} = \frac{(\Delta f + d)\tan\vartheta^{(2)} \sin\beta}{(\Delta f + d)\tan\vartheta^{(2)} \cos\beta - \Delta f \tan\vartheta^{(1)}}, \tag{S10}$$

where $\vec{\rho}_i$ are the positions of the virtual sources. The dependency described by Eq. S10, for (1)=hBN and (2)=graphene, is plotted in Fig. S3c. It exhibits a saturation behaviour at large rotation angles $\beta$. The simulated CBED patterns at different angle of rotation $\beta$ (shown in Fig. S3d) confirm that the tilt of the fringes $\mu$ changes fast only at small $\beta$.

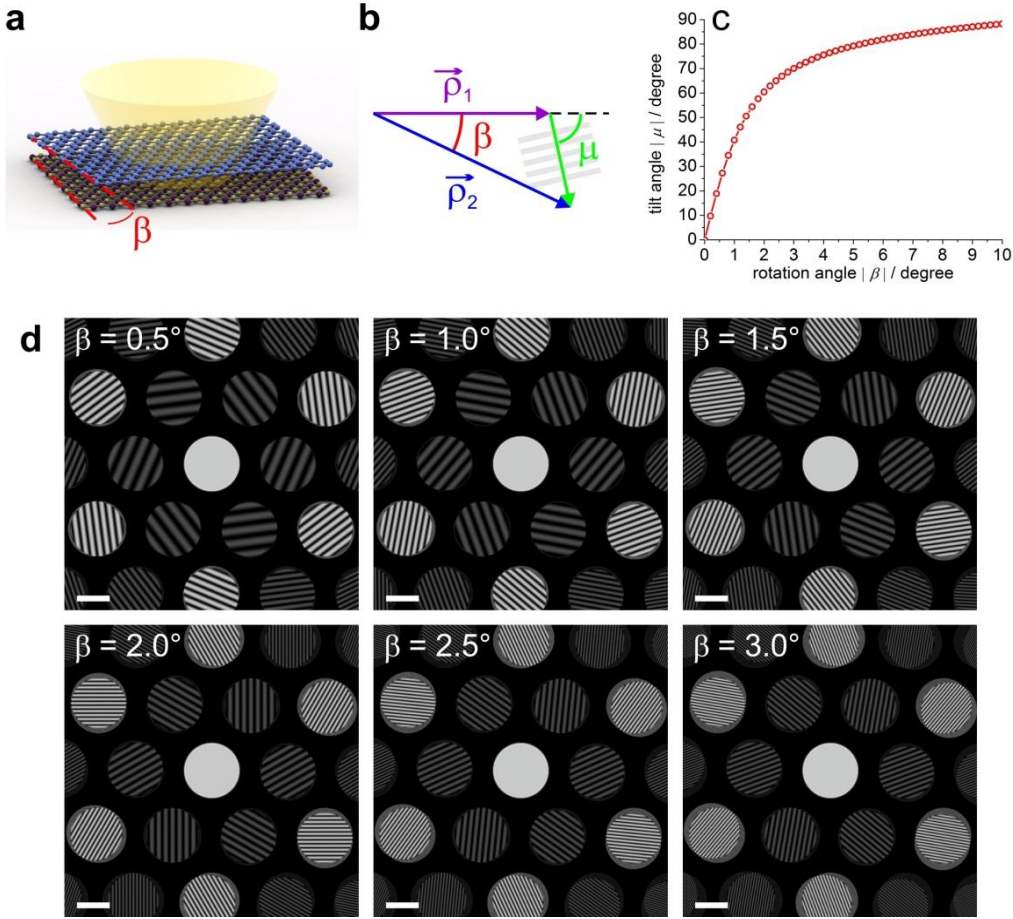

Fig. S3. Simulated CBED patterns of graphene-hBN samples where the graphene layer is rotated. (a) Schematic arrangement of 2D crystals and the incident electron beam. (b) Arrangement of the vectors in the virtual source plane. (c) Plot of fringes tilt angle $\mu$ as a function of the rotation angle $\beta$, according to Eq. S10. (d) Simulated CBED patterns of graphene-hBN samples where the graphene layer is rotated by $\beta$. For these simulations $\Delta f = -3$ μm, the distance between the layers is 3.35 Å, the imaged area is about 50 nm in diameter, the number of pixels is 512 × 512 and $\Delta K = 3.51\times10^7$ m$^{-1}$. The scale bars correspond to 2 nm$^{-1}$.

# 5. Phase shift caused by out-of- and in-plane displacements

## 5.1 Phase shift caused by an out-of-plane displacement

Wavefront scattered by an atom at $\vec{r} = (x, y, z)$ is given by Eq. S2:

$$U(K_x, K_y) \sim \exp(ikR)\exp(\pm ikr)\exp[-i(xK_x + yK_y)]\exp(-izK_z),$$

where "$-$" is for underfocus CBED ($\Delta f < 0$) and "$+$" is for overfocus CBED ($\Delta f > 0$), respectively.

For $\Delta f < 0$, wavefronts scattered by atoms positioned at $\vec{r}_1 = (0, 0, -|\Delta f|)$, $r_1 = |\Delta f|$ and $\vec{r}_2 = (0, 0, -|\Delta f| + \Delta z)$, $r_2 = |-|\Delta f| + \Delta z|$, are given by:

$$U_1(K_x, K_y) \sim \exp(ikR)\exp(-ik|\Delta f|)\exp[iK_z|\Delta f|]$$
$$U_2(K_x, K_y) \sim \exp(ikR)\exp[ik(-|\Delta f| + \Delta z)]\exp[-iK_z(-|\Delta f| + \Delta z)].$$

The corresponding phases of the wavefronts are:

$$\varphi_1 = kR - k|\Delta f| + \Delta f K_z$$
$$\varphi_2 = kR + k(-|\Delta f| + \Delta z) - (-|\Delta f| + \Delta z)K_z,$$

and the phase difference is given by:

$$\Delta\varphi_z = \varphi_2 - \varphi_1 = \Delta z(k - K_z) = \Delta z \frac{2\pi}{\lambda}(1 - \cos\vartheta),$$

where $K_z = \frac{2\pi}{\lambda}\cos\vartheta$. Using the relation $\frac{2\pi}{\lambda}\sin\vartheta = \frac{2\pi}{a}$ or $\cos\vartheta = \sqrt{1-\left(\frac{\lambda}{a}\right)^2} \approx 1 - \frac{1}{2}\left(\frac{\lambda}{a}\right)^2$,

where $a$ is the lattice constant, we obtain the phase difference:

$$\Delta\varphi_z = \Delta z \frac{2\pi}{\lambda}(1 - \cos\vartheta) \approx \Delta z \frac{2\pi}{\lambda}\frac{1}{2}\left(\frac{\lambda}{a}\right)^2 = \Delta z \frac{\pi\lambda}{a^2}. \tag{S11}$$

The phase shift $\Delta\varphi_z$ can be obtained by averaging the phase distributions reconstructed for individual CBED spots of the same order. The out-of-plane atomic displacements $\Delta z$ are then calculated from the obtained phase shift by applying Eq. S11.

## 5.2 Phase shift caused by an in-plane displacement

The wavefront scattered by an atom at $\vec{r} = (x, y, z)$ is given by Eq. S2:

$$U(K_x, K_y) \sim \exp(ikR)\exp(\pm ikr)\exp[-i(xK_x + yK_y)]\exp(-izK_z),$$

where "$-$" is for underfocus CBED ($\Delta f < 0$) and "$+$" is for overfocus CBED ($\Delta f > 0$), respectively.

For $\Delta f < 0$, wavefronts scattered by atoms positioned at $\vec{r}_1 = (0, 0, -|\Delta f|)$ and $\vec{r}_2 = (a + \Delta x, 0, -|\Delta f|)$, where $r_1 = |\Delta f|$ and $r_2 = \sqrt{(\Delta f)^2 + (a + \Delta x)^2} \approx |\Delta f|$, and $a$ is the lattice period, are given by:

$$U_1(K_x, K_y) \sim \exp(ikR)\exp(-ik|\Delta f|)\exp[iK_z|\Delta f|]$$
$$U_2(K_x, K_y) \sim \exp(ikR)\exp(-ik|\Delta f|)\exp[-iK_x(a + \Delta x)]\exp(iK_z|\Delta f|).$$

The corresponding phases of the wavefronts are:

$$\varphi_1(K_x, K_y) = kR - k|\Delta f| + K_z|\Delta f|$$
$$\varphi_2(K_x, K_y) = kR - k|\Delta f| - K_x(a + \Delta x) + K_z|\Delta f|,$$

and the phase difference is given by:

$$\Delta\varphi_x(K_x, K_y) = \varphi_2(K_x, K_y) - \varphi_1(K_x, K_y) = -K_x(a + \Delta x).$$

When $\Delta x = 0$, the phase shift $\Delta\varphi = 2\pi$, and we obtain:

$$\Delta\varphi_x(K_x, K_y) = -K_x^{(1)} a = 2\pi,$$

which corresponds to the position of the *n*-th-order diffraction peak

$$K_x^{(n)} = n\frac{2\pi}{a}.$$

The phase shift due to a lateral shift $\Delta x$ is given by

$$\Delta\varphi_x(K_x, K_y) = -K_x \Delta x \tag{S12}$$

which is an odd function of $K_x$. Thus, for $\Delta x \neq 0$ there will be additional phase shift in CBED spots (-1010) and (10-10), of opposite sign. Therefore, the phase shifts $\Delta\varphi_x(K_x, K_y)$ can be obtained by calculating the difference between the reconstructed phases of the opposite CBED spots and dividing the result by 2. The lateral atomic shifts $\Delta x$ are then calculated from the obtained phase shift by applying Eq. S12.

# 6. Simulated CBED patterns

## 6.1 Simulated CBED patterns with in-plane ripples at different rotation angles between the layers

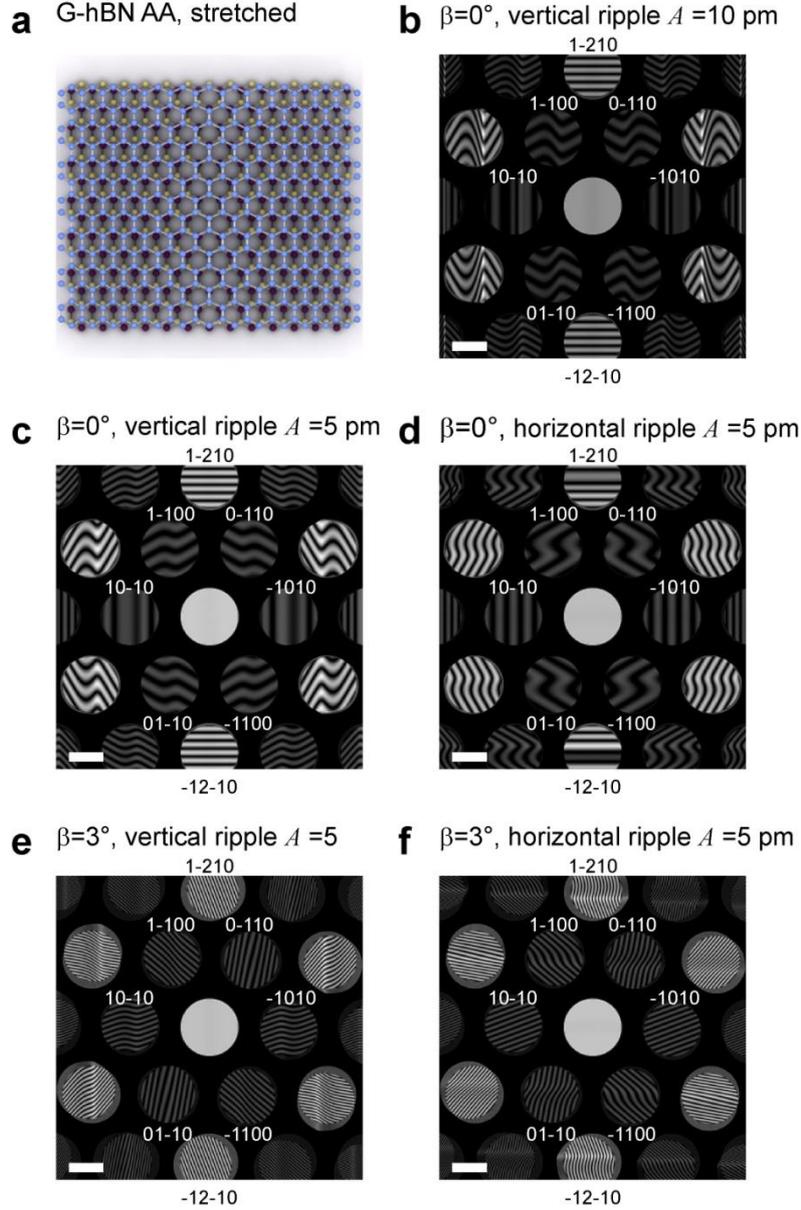

Fig. S4. Simulated CBED patterns of bilayer graphene-hBN samples where the graphene layer is rotated by angle $\beta$ and an in-plane ripple in graphene is introduced. (a) Sketch of top view on graphene-hBN sample with an in-plane stretched ripple in graphene. (b) Simulated CBED pattern of the sample, where the elongation of the $x$-component of the lattice period in graphene between $n$ and $n+1$ atoms is given by $\Delta x = A \exp\left[-\frac{(na_0)^2}{2\sigma_R^2}\right]$, $A = 10$ pm, $a_0 = 1.42$ Å, $\sigma_R = 4$ nm. (c) Vertical in-plane ripple, $A = 5$ pm and $\beta = 0°$. (d) Horizontal in-plane ripple, $A = 5$ pm and $\beta = 0°$. (e) Vertical in-plane ripple, $A = 5$ pm and $\beta = 3°$. (f) Horizontal in-plane ripple, $A = 5$ pm and $\beta = 3°$. For these simulations $\Delta f = -3$ μm, the distance between the layers is 3.35 Å, the imaged area is about 50 nm in diameter, the number of pixels is 512 × 512 and $\Delta K = 3.51 \times 10^7$ m$^{-1}$. The scale bars in (b) – (f) correspond to 2 nm$^{-1}$.

## 6.2 Simulated CBED patterns with out-of-plane ripples at different rotation angles between the layers

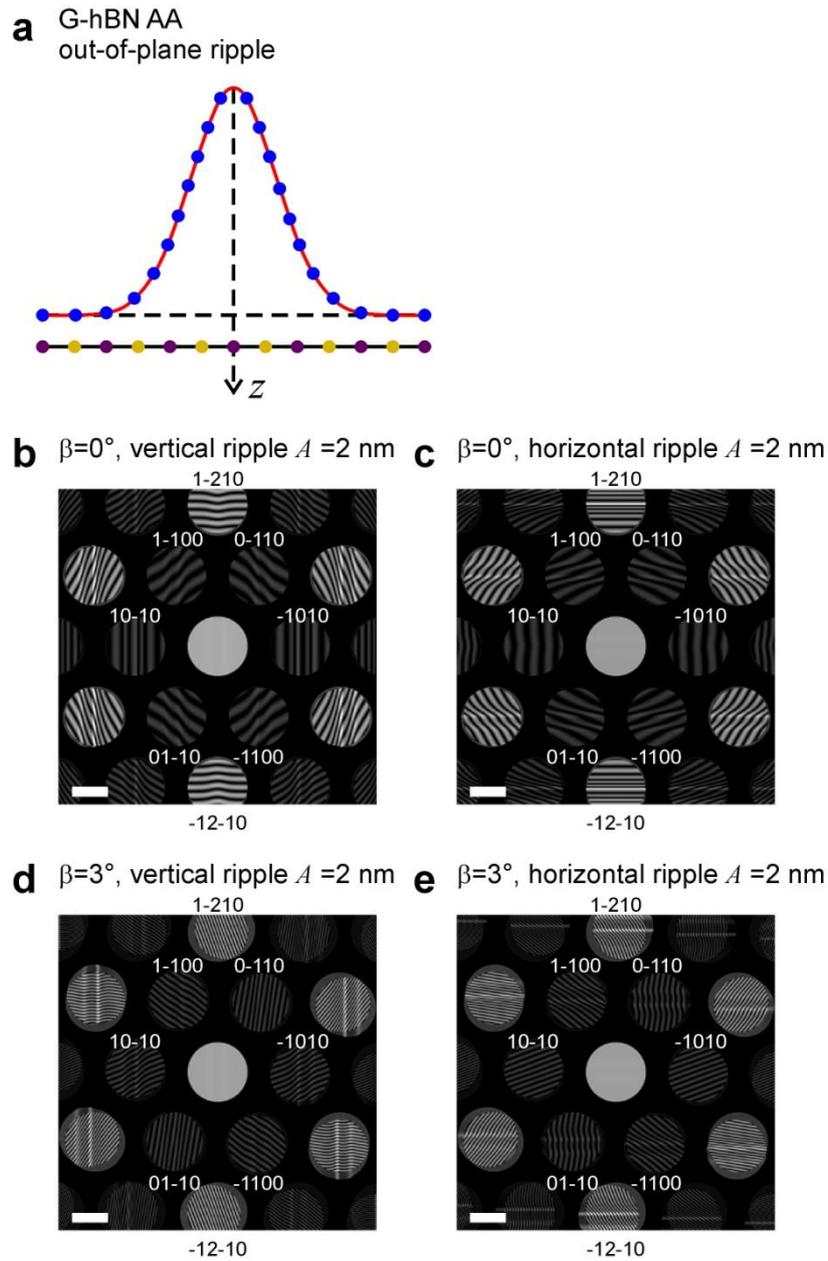

Fig. S5. Simulated CBED patterns of bilayer graphene-hBN samples where the graphene layer is rotated by angle $\beta$ and an out-of-plane ripple in graphene is introduced. (a) Side view on graphene-hBN sample with an out-of-plane ripple in graphene. The distance between the atoms is kept as in pristine graphene, so no strain (apart of that due to bending) is involved. (b) The related simulated CBED pattern of the sample where the atomic $z$-positions in graphene are shifted by $\Delta z = -A \exp\left(-\dfrac{(x-x_0)^2}{2\sigma_R^2}\right)$, $A = 2$ nm, $\sigma_R = 5$ nm, $x_0 = 0$. (c) Horizontal out-of-plane ripple, $A = 2$ nm and $\beta = 0°$. (d) Vertical out-of-plane ripple, $A = 2$ pm and $\beta = 3°$. (e) Horizontal out-of-plane ripple, $A = 2$ pm and $\beta = 3°$. For these simulations $\Delta f = -3$ μm, the distance between the layers is 3.35 Å, the imaged area is about 50 nm in diameter, the number of pixels is 512 × 512 and $\Delta K = 3.51 \times 10^7$ m$^{-1}$. The scale bars in (b) – (e) correspond to 2 nm$^{-1}$.

# 7. Reconstruction of an individual CBED spot as off-axis holograms

## 7. 1 Steps of reconstruction

The intensity distribution in a bilayer CBED spot can be described as

$$I = |U_1|^2 + |U_2|^2 + U_1^* U_2 + U_1 U_2^*,$$

where $U_1$ and $U_2$ are the waves diffracted from the two layers, respectively. For simplicity we consider non-rotated layers and we select coordinates in the detector plane $(X,Y)$ so that the fringes are distributed along the $X$-axis. The wave scattered by the first layer is given by:

$$U_1(X) = U_{1,0}(X) \exp\left( \frac{2\pi i}{\lambda} X \sin \vartheta^{(1)} + i\varphi_1(X) \right)$$

and the wave scattered by the second layer is described by

$$U_2(X) = U_{2,0}(X) \exp\left( \frac{2\pi i}{\lambda} X \sin \vartheta^{(2)} + i\varphi_2(X) \right),$$

where $\vartheta^{(1)}$ and $\vartheta^{(2)}$ are the scattering angles corresponding to the first-order diffraction peaks, and $\varphi_1(X)$ and $\varphi_2(X)$ are the phase shifts because of the lattices corrugation. The resulting interference pattern is given by

$$I(X) = |U_{1,0}(X)|^2 + |U_{2,0}(X)|^2 + U_{1,0}^*(X) U_{2,0}(X) \exp\left\{ \frac{2\pi i}{\lambda} X \left( \sin \vartheta^{(2)} - \sin \vartheta^{(1)} \right) + i\Delta\varphi(X) \right\} + c.c.$$

where we introduced the phase difference $\Delta\varphi(X) = \varphi_2(X) - \varphi_1(X)$.

The reconstruction steps are:

1. A CBED spot is selected, placed into the centre and multiplied with an apodization function so that all signal around the spot is set to zero.

2. Fourier spectrum of the CBED spot is calculated. Two sidebands are obtained at the frequencies $\nu = \pm \frac{1}{\lambda} \left( \sin \vartheta^{(2)} - \sin \vartheta^{(1)} \right)$. For a known angle of the relative rotation between the two perfect lattices, the positions of the sidebands in the Fourier spectrum can be easily calculated.

3. One side-band is selected and re-positioned to the centre.

4. The inverse Fourier transform of the resulting spectrum gives a complex-valued distribution where the phase is $\Delta\varphi(X)$.

## 7.2 Example of reconstruction of CBED spots as off-axis holograms

In this section we explain in more detail the reconstruction steps using the example of CBED pattern of bilayer graphene-hBN with an out-of-plane bubble in graphene, shown in Fig. S6a. hBN layer was rotated by $\beta = 4°$ relatively to the graphene layer. Because we assume the following ordering of the layers along optical axis: graphene layer before the hBN layer, the $z$-positions of atoms in graphene are shifted as $\Delta z = -A \exp\left(-\frac{x^2 + y^2}{2\sigma_B^2}\right)$, where $A = 2$ nm and $\sigma_B = 2$ nm. Note that the presence of the bubble is more noticeable in the second- and higher-order CBED spots, Fig. S6a.

The $z$-positions of atoms in the graphene bubble are shifted as $\Delta z = -A \exp\left(-\frac{x^2 + y^2}{2\sigma_B^2}\right)$, which can be thought as a phase shifting object. One can notice that such a phase-shifting distribution is analogous to transmission function of a lens and the intensity distribution in the far-field thus is expected to have a maxima at the centre of the lens because of the focusing effect. This explains why the intensity in CBED spots has maxima at the position of the bubble as observed in the CBED patterns of the graphene layer alone, Fig. 3b (main text). The phase distribution, in turn, follows the distribution $\left(-\frac{x^2 + y^2}{2\sigma_B^2}\right)$, with a minimum at the position of the bubble.

The reconstruction procedure is performed as described above, and it is sketched in Fig. S6.

(i) Positions of CBED spots of individual layers are calculated from the known lattice constants of the individual layers and the angle of the relative rotation between the layers.

(ii) A bilayer CBED spot is selected and re-positioned to the centre of the image, Fig. S6a – b.

(iii) Fourier transform of the resulting distribution is calculated. In the obtained Fourier spectrum, one sideband is selected (at the position of the theoretically calculated position of the sideband) and shifted to the centre, as shown in Fig. S6c. The inverse Fourier transform of the resulting spectrum is a complex-valued distribution which provides amplitude and phase distribution reconstructed for the selected CBED spot, Fig. S6d.

(iv) Steps (ii) and (iii) are repeated for each of the six CBED spots of the same order, the obtained distributions are shown in Fig. S6e – f.

(v) For further analysis, the layer of interest identified. For example, if deformations are in the graphene layer, then the reconstructed complex-valued distributions are extracted at the regions with centres corresponding to CBED spots of graphene layer. These reconstructed complex-valued distributions are then the subject to summation or subtraction of phases depending on what displacement needs to be extracted, out-of-plane or in-plane, respectively.

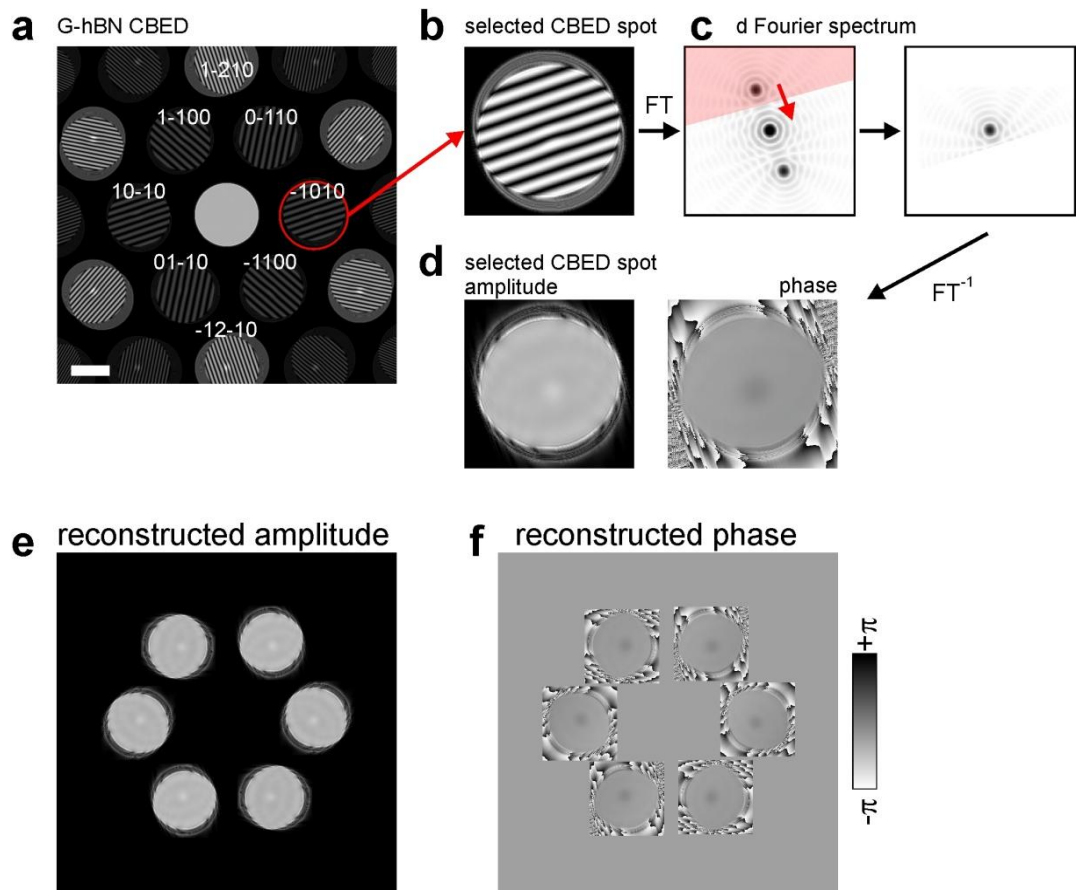

Fig. S6. Reconstruction of CBED spots as off-axis holograms.

(a) Simulated CBED patterns of graphene with a bubble. For these simulations $\Delta f = -2$ μm, the distance between the layers is 3.35 Å, the imaged area is about 28 nm in diameter, the number of pixels is 2048 × 2048 and $\Delta K = 8.77 \times 10^6$ m$^{-1}$. The scalebar corresponds to 2 nm$^{-1}$.

(b) Magnified selected spot (-1010) (circled in (a)) where irregularities of the fringe pattern can be seen.

(c) Fourier transform of the CBED spot intensity distribution is calculated, its amplitude distribution exhibits the zero-order and two sidebands. The area marked by red is selected and shifted so that the sideband peak is in the centre.

(d) Inverse Fourier transform of the selected sideband gives the complex-valued distribution of the wavefront in the detector plane.

(e) Amplitude and (f) phase distributions reconstructed from the six first-order CBED spots.

## 7.3 Example of reconstruction of a mixed in-plane and out of-plane displacement

Fig. S7 shows reconstruction of CBED pattern of graphene-hBN bilayer, when both $\Delta x$ and $\Delta z$ atomic shifts occur in the graphene layer. The simulated CBED pattern is shown in Fig. S7a. The difference between the phases of the wavefronts scattered of the graphene layer with and without atomic displacement is shown in Fig. S7b. Fig. S7c and d show the reconstructed amplitude and the phase distributions obtained by applying off-axis holographic reconstruction routine for (-1010) CBED spot. In these reconstructed distributions, a blurring and the vertical interference fringes are observed in the region of $x \approx 0$. These effects can be explained by the diffraction on knife-edge. Because of infinite coherence assumed in the simulation - Fresnel fringes appear and the transition region is blurred by about $2.4\sqrt{\lambda \Delta f / 2}$. The complex-valued distribution at a selected CBED spot $o'(x', y')$ can be described by Fresnel diffraction of the sample distribution $o(x, y)$:

$$o'(x', y') = \iint o(x, y) \exp\left[-\frac{i\pi}{\lambda \Delta f}(x - x')^2 + (y - y')^2\right] dx \, dy,$$

which is a convolution of sample distribution $o(x, y)$ with the spherical wave distribution. Therefore, $o(x, y)$ can be obtained by performing backward propagation of reconstructed $o'(x', y')$ ("deconvolution"):

$$o(x, y) = \iint o'(x', y') \exp\left[\frac{i\pi}{\lambda \Delta f}(x - x')^2 + (y - y')^2\right] dx' \, dy'. \tag{S13}$$

Numerical algorithms for the wavefront propagation can be found in (2). Fig. S7e and f show amplitude and phase of the complex-valued distribution obtained after applying backward propagation of the reconstructed wavefront for (-1010) CBED spot. To avoid diffraction effects due to backward propagation procedure, edges of the CBED spot were blurred by apodization function. It is apparent that a sharp edge at $x = 0$ is now reconstructed in the phase distribution, as shown in Fig. S7f.

Each of the six first-order CBED spots was analyzed in the same manner, as described above: wavefront was reconstructed by applying off-axis holographic reconstruction routine and the resulting wavefront was backward propagated by applying integral transformation given by Eq. S13. The atomic displacements $\Delta z$ and $\Delta x$ are then calculated by applying Eq. 2 and 3 (main text) and the resulting distributions are shown in Fig. S7g – h.

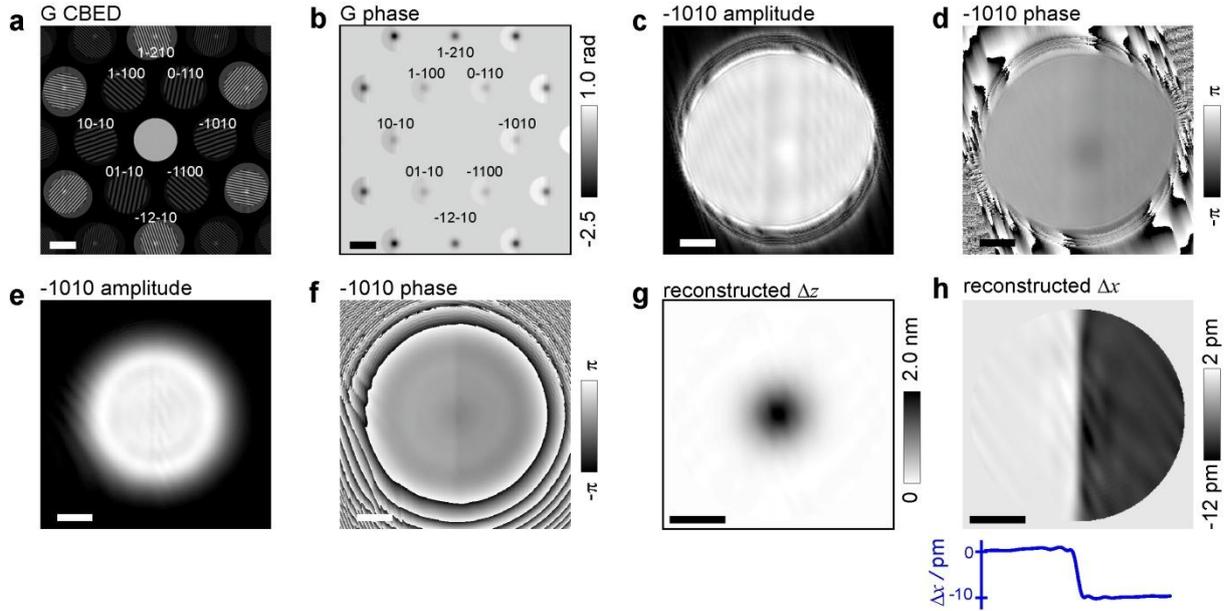

Fig. S7. Simulated CBED patterns for graphene and graphene-hBN bilayer heterostructures where graphene lattice is deformed.

(a) Simulated CBED pattern where the atoms positioned at $x>0$ are displaced by $\Delta x = -10$ pm, and the atomic $z$-positions are shifted by $\Delta z = -A_B \exp\left(-\dfrac{x^2+y^2}{2\sigma_B^2}\right)$, $A_B = 2$ nm, $\sigma_B = 2$ nm. For these simulations $\Delta f = -2$ μm, the distance between the layers is 3.35 Å, the imaged area is about 28 nm in diameter, the number of pixels is 2048 × 2048 and $\Delta K = 8.77 \times 10^6$ m$^{-1}$.

(b) The difference of the phases of the wavefronts scattered of the graphene layer with and without atomic displacement.
(c) Amplitude of the reconstructed wavefront at (-1010) CBED spot.
(d) Phase of the reconstructed wavefront at (-1010) CBED spot.
(c) Amplitude of the reconstructed complex-valued distribution from (-1010) CBED spot after deconvolution.
(d) Phase of the reconstructed complex-valued distribution from (-1010) CBED spot after deconvolution.
(g) Reconstructed distribution of the atomic out-of-plane displacement, $\Delta z$, in the graphene layer.
(h) Reconstructed distribution of the atomic in-plane displacement, $\Delta x$, in the graphene layer.

The scale bars in (a) and (b) correspond to 2 nm$^{-1}$. The scale bars in (c) – (h) correspond to 5 nm.

## 7.4 Note on reconstruction of exact distance between the layers

The period $T$ and the tilt of the fringes $\mu$ in a CBED spot interference pattern are unambiguously defined through such parameters as lattices periods (and associated scattering angles $\vartheta^{(1)}$ $\vartheta^{(2)}$), the probing electron beam energy $\lambda$, the sample $z$-position $\Delta f$, the relative rotation between the layers $\beta$, and the distance between the layers $d$. $\Delta f$ is approximately known from experiment and $d$ is typically unknown. From Equations S9 and S10, both, $\Delta f$ and the distance between the layers $d$, can be evaluated from the measured period $T$ and the tilt of the fringes $\mu$, provided that $\vartheta^{(1)}$, $\vartheta^{(2)}$ and $\beta$ are known:

$$\Delta f \approx \frac{\Delta \rho}{\sqrt{\xi}},$$

$$d \approx \frac{\Delta \rho}{\sqrt{\xi}} \frac{\tan \vartheta^{(2)} \sin \beta - \tan \mu \tan \vartheta^{(2)} \cos \beta + \tan \mu \tan \vartheta^{(1)}}{\tan \mu \tan \vartheta^{(2)} \cos \beta - \tan \vartheta^{(2)} \sin \beta},$$

where

$$\xi = \tan^2 \vartheta^{(1)} + \tan^2 \vartheta^{(2)} - 2 \tan \vartheta^{(1)} \tan \vartheta^{(2)} \cos \beta.$$

From Eq. S8, for $\beta = 0$:

$$d = \frac{\Delta f \left( \tan \vartheta^{(1)} - \tan \vartheta^{(2)} \right) - \Delta \rho}{\tan \vartheta^{(2)}}.$$

The precise positions of Fourier sideband peaks calculated from the period and the tilt of the interference fringes are not integer numbers when measured in pixels. For example, for CBED pattern of graphene-hBN bilayer ($\Delta f = 2\,\mu m$, 2048 × 2048 pixels, $\Delta K = 8.77 \times 10^6$ m$^{-1}$), the calculated precise positions of the sideband peaks in the Fourier spectra of the first-order CBED spots, for $d = 0$ and $d = 6$ Å are summarized in Table 1. The sideband peaks positions measured in integer pixels, however, are the same for both distances, $d = 0$ and $d = 6$ Å, as evident from Table 2. Thus, from the sideband peaks positions measured in integer pixels one cannot derive the exact distance between the layers.

| $d = 0$ | -1010 | 0-110 | 1-100 | 10-10 | 01-10 | -1100 |
|---|---|---|---|---|---|---|
| X / px | 1009.63 | 1016.81 | 1031.19 | 1038.37 | 1031.19 | 1016.81 |
| Y / px | 1024 | 1011.55 | 1011.55 | 1024 | 1036.45 | 1036.45 |

| $d = 6$ Å | -1010 | 0-110 | 1-100 | 10-10 | 01-10 | -1100 |
|---|---|---|---|---|---|---|
| X / px | 1009.83 | 1016.92 | 1031.08 | 1038.17 | 1031.008 | 1016.92 |
| Y / px | 1024 | 1011.73 | 1011.73 | 1024 | 1036.27 | 1036.27 |

Table 1

|  | -1010 | 0-110 | 1-100 | 10-10 | 01-10 | -1100 |
|---|---|---|---|---|---|---|
| X / px | 1010 | 1017 | 1031 | 1038 | 1031 | 1012 |
| Y / px | 1024 | 1012 | 1012 | 1024 | 1036 | 1036 |

Table 2

Nevertheless, the information about the interlayer distance is stored in the entire distribution of interference pattern and can be readily reconstructed by the off-axis holographic approach. Fig. S8a shows four CBED patterns of graphene-hBN layers at interlayer distances $d = 0, 3, 6$ and $10$ Å, which visually exhibit the same distributions. Fig. S8b shows the profiles of the reconstructed interlayer distances. The artifact low-frequency signal in the reconstructed interlayer distances in Fig. S8b can be explained by the fact that the exact sideband positions were replaced by the sideband positions in integer pixels. The distribution of the interlayer distance reconstructed from CBED at $d = 0$ provides the off-set which should be subtracted from the distances reconstructed from CBED at $d \neq 0$. This way, correct values of interlayer distances can obtained. For typical experimental parameters (as for example in the simulated example here: $\Delta f = 2\,\mu\text{m}$, 2048 × 2048 pixels, $\Delta K = 8.77 \times 10^6 \text{ m}^{-1}$) the offset value is very small and can be neglected.

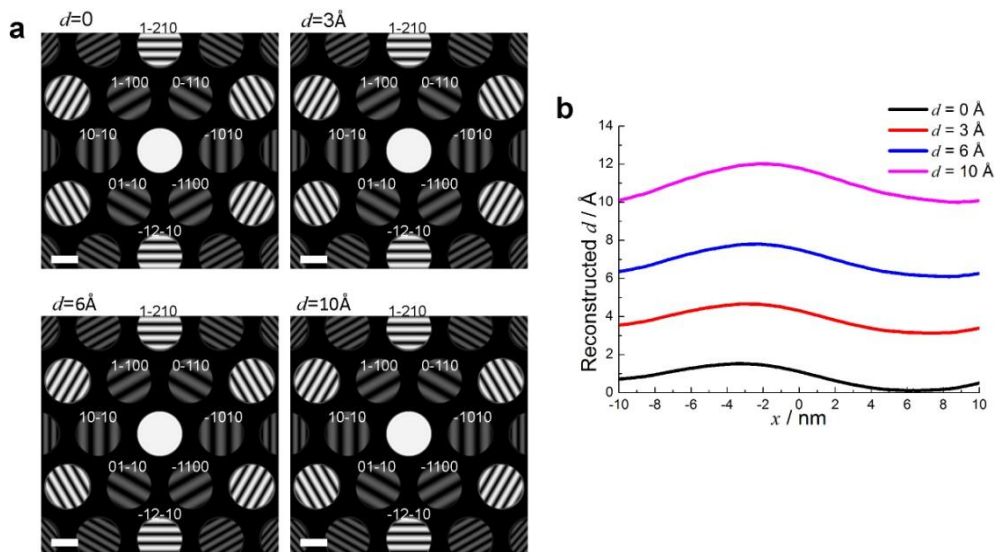

Fig. S8. Reconstruction of distances between the layers. (a) Simulated CBED patterns for graphene-hBN bilayer heterostructures at different interlayer distances. (b) Profiles of the reconstructed interlayer distances across the sample.

## 7.5 Spatial resolution

The spatial resolution provided by the off-axis holographic reconstruction procedure is defined by the size of the mask applied to cut out a selected sideband in the Fourier domain. Such a mask only crops the spectrum on one side, as shown in the example in Fig. S6b. The resulting resolution approximately equals to half of the period of fringes. The period of fringes, when transformed from *K*-domain to sample domain, is given by:

$$T = T_K \frac{\lambda}{2\pi} \Delta f = \frac{\lambda}{\Delta \rho} \Delta f,$$

where $T_K$ is in the period of fringes in *K*-domain as given by Eq. S9, and $\Delta \rho$ is the distance between the virtual sources as defined by Eq. S8.

For example, the spatial resolution of the reconstruction shown in Fig. 5 (main text) is about 2 nm ($\Delta f = -3 \ \mu m, \ \Delta \rho = 3.27 \ nm$). It should be noted that the resolution can be increased by reconstructing atomic displacements from the second- or higher-order CBED spots, for which the distance between the virtual sources $\Delta \rho$ is higher.

## 8. TEM

TEM CBED imaging was performed with a probe side aberration corrected Titan ChemiSTEM operated at 80 kV with a small convergence angle and a probe current of 110 pA. The images were recorded with a 16 bit intensity dynamic range detecting system, without using a beam spot, so that the intensity in all CBED diffraction spots is available. Each image was the average of 10 identical acquisitions, with a 1 s acquisition time.

# 9. Experimental CBED patterns at different Δf

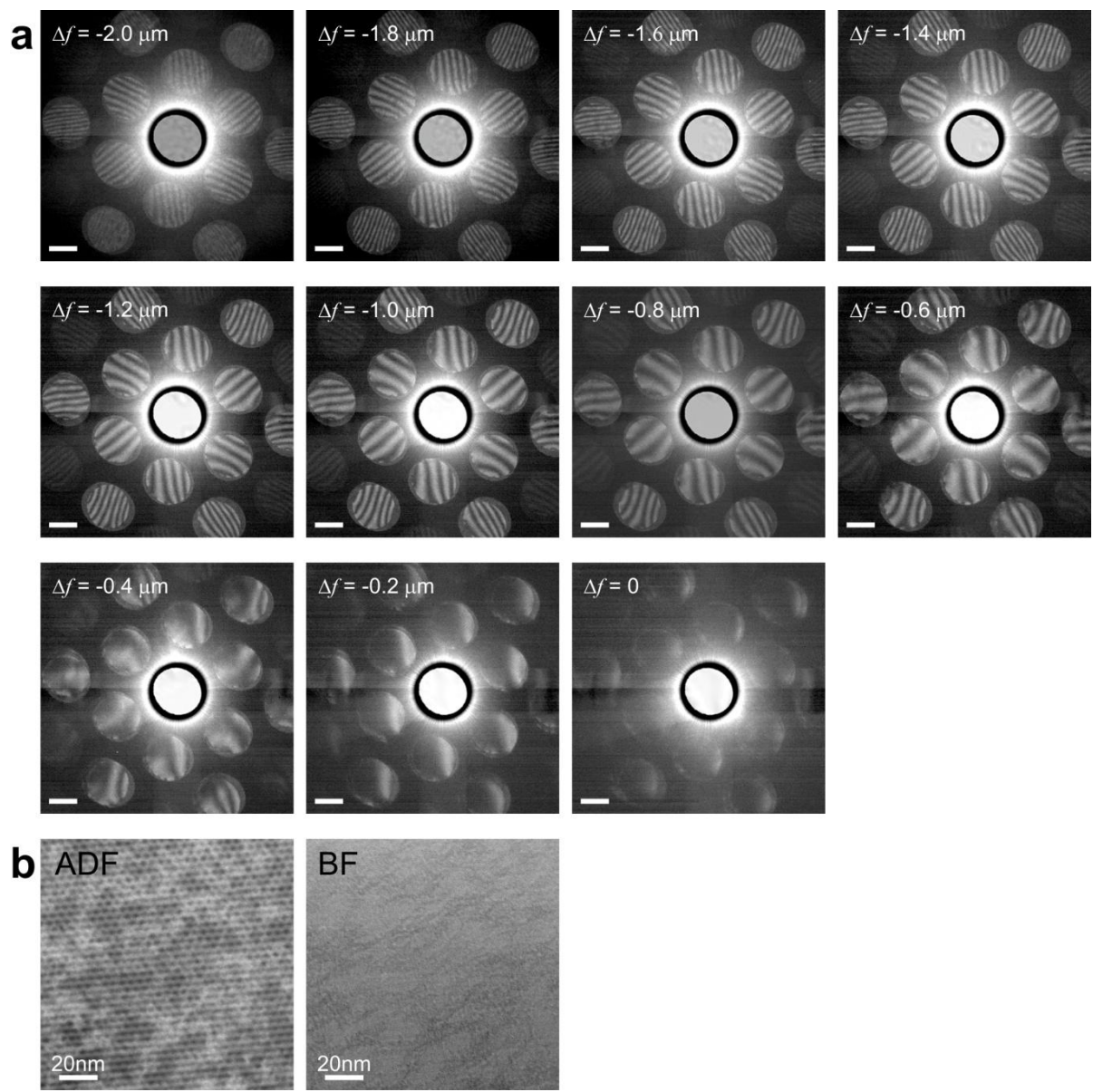

Fig. S9. (a) Experimental CBED patterns of graphene-hBN bilayer obtained at different $\Delta f$. The scalebar is 2 nm$^{-1}$. The intensity of the central spot is reduced by factor 0.001. (b) Annular dark field (ADF) and bright field (BF) images of the same sample area.

## 10. CBED pattern of an edge

CBED patterns shown in Fig. S10 demonstrate how the interference pattern disappears when part of one of the layers is missing. Note that in the simulated images, where infinite coherence is assumed, the diffraction causes the interference pattern to be continued into the regions where there is no second layer, Fig. S10b. In the experimental images, due to partial coherence of electron waves, the edge of the layer appears almost sharp without noticeable diffraction effects.

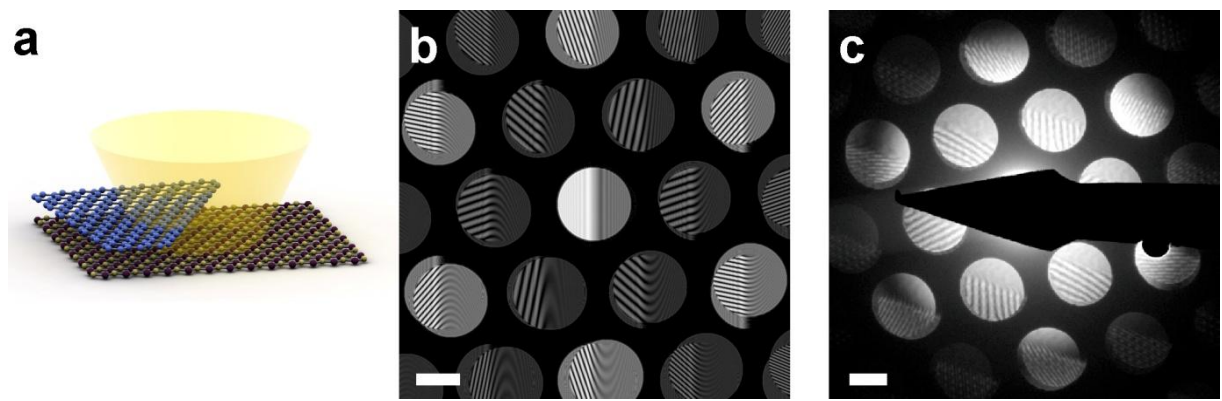

Fig. S10. Simulated and experimental CBED pattern of a graphene-hBN sample at the edge of one layer. (a) Illustration of experimental arrangement. (b) Simulated CBED pattern. For these simulations the distance between the layers is 3.35 Å, $\Delta f = -2$ μm, and the imaged area is about 28 nm in diameter. (c) Experimental CBED pattern. The scale bars in (c) and (d) correspond to 2 nm$^{-1}$.

## 11. Origin of rippling

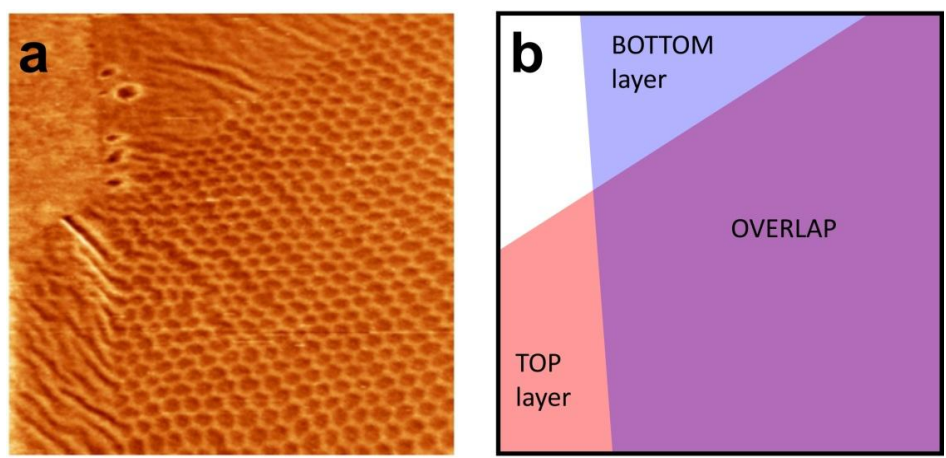

Fig. S11. (a) Topographic image of a monolayer-hBN (mhBN) on graphene sample on SiO$_2$. The size of the scan is 400 nm. Z-contrast (from black to white): 6 nm. (b). Cartoon illustrating 4 regions formed by the overlapping layers in (a).

Even though we only extract the relative information between the layers – we still can know the particular layer which contains the ripple due to specific sample preparation procedure. Here we used the pick and lift method. This involves exfoliating graphene onto a PMMA substrate and the hBN onto a SiO$_2$ substrate. Then, the PMMA is used as a membrane to suspend the graphene over the hBN crystal as the two are brought into contact. The crystals adhere and the membrane is lifted again with the two crystals attached. The two crystals are then positioned over a hole in the TEM grid and brought into contact with it. The PMMA is then removed with acetone.

We anticipate that the PMMA is ultimately responsible for the rippling of the crystal exfoliated onto it, perhaps as a result of uneven drying when the polymer layer is spin-coated. This is corroborated by the sample in the figure above. Fig. S11 shows topography image of monolayer-hBN (mhBN) on graphene sample on $SiO_2$. Four distinct regions are present in the image; $SiO_2$ only (top left), graphene on $SiO_2$ (top right), mhBN on $SiO_2$ (bottom left), and mhBN on graphene on $SiO_2$ (bottom right). The angular dependant moiré pattern is clearly visible in the overlap regions. However, also observable is that there is a large amount of rippling in the mhBN flake. This suggests the PMMA introduces (or allows) significant amounts of wrinkling into the crystal in contact with it.

## 12. Alternative methods of observing defects in graphene/hBN heterostructures

CBED is uniquely positioned for the observation of atomic scale defects in van der Waals heterostructures. However, here we made an attempt of independent verification of our CBED observation. We used AFM to investigate the same area we observed in TEM. We would like to point out that AFM would be sensitive only to certain types of defects (like out-of-plane ripples) and would be completely insensitive to such defects as in-plane ripples, local strain, etc – which are still observable by CBED.

The AFM image of approximately the same region as we observed in TEM is presented in Fig. S12. We would like to point out that it is very difficult to identify the exact region we observed in TEM: our membranes are few micrometre in diameter and the region we observe in CBED is few tens of nanometres. The AFM image demonstrates surface with high roughness (~1nm), which is probably due to residual contamination acquired through the transfer process (wet transfer is usually used for transfer of van der Waals heterostructures on TEM grids) and a contamination layer due to exposure to the electron beam.

Also, we observed linear defects, which might be identified as ripples. Their height is a few nanometres, similar to the ripples heights reconstructed from CBED patterns. We would like to stress, that the information obtained by AFM cannot allow us to be more conclusive about the nature of the defects. It is only possible to say that the AFM observation is in agreement with our CBED finding. Note, our AFM study doesn't provide any information on the width of the ripple as an AFM image is always a convolution between the shape of the ripple and the shape of the AFM tip.

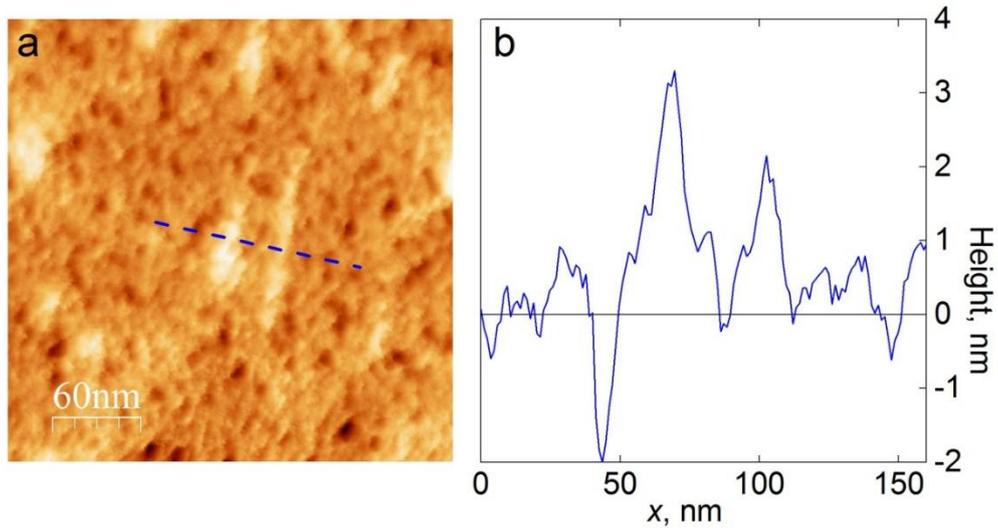

Fig. S12. AFM investigation of the atomic scale defects in van der Waals heterostructures. (a) Topographic image of the same graphene-hBN sample and approximately the same spot as observed in Fig. 5 in the main text. Linear defects which can be ripples are visible. (b) Cross-section of (a) along the dashed blue line.

## References

1. Latychevskaia T & Fink H-W (2007) Solution to the twin image problem in holography. *Phys. Rev. Lett.* 98(23):233901.
2. Latychevskaia T & Fink H-W (2015) Practical algorithms for simulation and reconstruction of digital in-line holograms. *Appl. Optics* 54(9):2424–2434.